\documentclass[useAMS,usenatbib,usegraphicx,referee]{mn2e}
\newcommand{\bugref}{\bibitem[\protect\citeauthoryear{dummy }{1893}]{dum}}

\title[Faraday Rotation Gradients in B1803+784]
{Surprising Evolution of the Parsec-scale Faraday Rotation Gradients in the Jet of the BL~Lac Object B1803+784}

\author[M. Mahmud, D. C. Gabuzda \& V. Bezrukovs]{M. Mahmud$^{1}$, D. C. Gabuzda $^{1}$ \& V. Bezrukovs$^{2}$ \\
$^{1}$Physics Department, University College Cork, Cork, Ireland \\
$^{2}$Applied Physics and Instrumentation Department, Cork Institute of Technology, Cork, Ireland}

\begin{document}

\date{Released 2009 Xxxxx XX}
\pagerange{\pageref{firstpage}--\pageref{lastpage}} \pubyear{2009}
\maketitle
\label{firstpage}
\begin{abstract}

Several multi-frequency polarization studies have shown the presence of 
systematic Faraday Rotation gradients across the parsec-scale jets of Active 
Galactic Nuclei (AGN), taken to be due to the systematic variation of the 
line-of-sight component of a helical magnetic ({\bf B}) field across the jet. 
Other studies have confirmed the presence and sense of these gradients in 
several sources, thus providing evidence that these gradients persist over 
time and over large distances from the core. However, we find surprising new 
evidence for a reversal in the direction of the Faraday Rotation gradient 
across the jet of B1803+784, for which multi-frequency polarization 
observations are available at four epochs. At our three epochs and the epoch 
of Zavala \& Taylor (2003), we observe transverse Rotation Measure (RM) 
gradients across the jet, consistent with the presence of a helical magnetic 
field wrapped around the jet. However, we also observe a ``flip'' in the 
direction of the gradient between June 2000 and August 2002. Although the 
origins of this phenomena are not entirely clear, possibly explanations 
include (i) the sense of rotation of the central 
supermassive black hole and accretion disc has remained the same, but the 
dominant magnetic pole facing the Earth has changed from North to South;
(ii) a change in the direction of the azimuthal {\bf B} field component as a 
result of torsional oscillations of the jet; and (iii) a change in the 
relative contributions to the observed rotation measures of the ``inner'' 
and ``outer'' helical fields in a magnetic-tower model. Although we cannot
entirely rule out the possibility that the observed changes in the 
RM distribution are associated instead with changes in the thermal-electron
distribution in the vicinity of the jet, we argue that this explanation
is unlikely. 
\end{abstract}

\section{Introduction}

BL~Lac objects are Active Galactic Nuclei (AGN), characterized by strong and 
variable polarization, rapid variability in luminosity, a featureless spectrum 
and weak optical line emission. The radio emission associated with BL~Lac 
objects is synchrotron emission, which can be linearly polarized up to about 
75\% in the optically thin (jet) region, and up to 10--15\% in the optically 
thick (core) region (Pacholczyk 1970). VLBI polarization observations of 
BL~Lac objects have shown a tendency for the polarization {\bf E} vectors in 
the parsec-scale jets to be aligned with the local jet direction, which 
implies that the corresponding {\bf B} field is transverse to the jet, because 
the jet is optically thin (Gabuzda, Pushkarev \& Cawthorne 2000). Although in 
the past, the dominance of the transverse {\bf B} field component was suggested 
to be the consequence of a `shock model' where a series of relativistic shocks 
compress and enhance the transverse {\bf B} field component (Laing 1980; 
Hughes, Aller \& Aller 1989), this seems an improbable explanation for the 
transverse fields detected in extended regions in the jets of some sources. 
Instead, a helical {\bf B} field associated with the jet, with the toroidal 
component dominating over the longitudinal component, would be a more 
plausible explanation (Lyutikov, Pariev \& Gabuzda 2005). In fact, systematic 
gradients in the Faraday rotation have been observed across the parsec-scale 
jets of a number of AGN, interpreted as reflecting the systematic change in the 
line-of-sight component of a toroidal or helical jet {\bf B} field across the 
jet (Asada et al. 2002; Gabuzda, Murray, Cronin 2004; Zavala \& Taylor 2005;
Gabuzda et al. 2008; Mahmud \& Gabuzda 2008; Asada et al. 2008a,b); such fields 
would come about in a natural way as a 
result of the ``winding up'' of an initial ``seed'' field by the rotation of 
the central accreting objects (e.g. Nakamura, Uchida \& Hirose 2001: Lovelace 
et al. 2002).

Faraday Rotation studies are crucial in determining the intrinsic {\bf B} 
field geometries associated with the jets. Faraday Rotation of the plane of 
linear  polarization occurs during the passage of an electromagnetic wave 
through a region with free electrons and a magnetic field with a non-zero 
component along the line-of-sight. The amount of rotation is proportional to 
the integral of the density of free electrons $n_{e}$ multiplied by the 
line-of-sight B field $B \cdot dl$, the square of the observing wavelength 
$\lambda^{2}$, and various physical constants; the coefficient of 
$\lambda^{2}$  is called the Rotation Measure (RM):
\begin{eqnarray}
           \Delta\chi\propto\lambda^{2}\int n_{e} B\cdot dl\equiv RM\lambda^{2}
\end{eqnarray}
The intrinsic polarization angle can be obtained from the following equation:
\begin{eqnarray}
           \chi_{obs} =  \chi_0 + RM \lambda^{2}
\end{eqnarray}
where $\chi_{obs}$ is the observed polarization angle and $\chi_0$ is the 
intrinsic polarization angle observed if no rotation occurred 
(Burn 1966). Simultaneous multifrequency 
observations thus allow the determination of the RM, as well as identifying 
the intrinsic polarization angles $\chi_0$.

B1803+784 has been studied using VLBI for nearly three decades. The predominant 
jet direction in centimetre-wavelength images is toward the West. The 
dominant jet {\bf B} field is perpendicular to the local jet direction 
essentially throughout the jet, from distances of less than 1~mas from the 
VLBI core (Gabuzda 1999, Lister 2001) to tens of mas from the core (Gabuzda \& 
Chernetskii 2003; Hallahan \& Gabuzda 2009); further, the {\bf  B} field
remains orthogonal even in extended regions and in the presence of appreciable 
bending of the jet.
Therefore, it seems most likely that this transverse jet {\bf B} field 
primarily represents the  toroidal component of the intrinsic {\bf B} field 
of the jet, rather than a series of transverse shocks (of course, this does
not rule out the possibility that some individual compact features may
be shocks). We have 
detected a transverse RM gradient across the VLBI jet; although it is
difficult to prove conclusively, combined with the observation of orthogonal
{\bf B} fields throughout the jet, even in the presence of appreciable
bending, this provides direct 
evidence that the jet has a helical {\bf B} field. Comparison of the 
gradients observed for several different epochs shows that the 
direction of the gradient changed sometime between June 2000 and August 
2002. We discuss the 
data demonstrating this unexpected change, as well as possible origins of the 
observed reversal of the RM gradient.

\section{Faraday-Rotation Observations and Reduction}
We consider here polarization data for B1803+784 obtained using the ten 25-m 
radio telescopes of the Very Long Baseline Array (VLBA) at four different 
epochs: 6 April 1997 (Gabuzda \& Chernetskii 2003), 27 June 2000 (Zavala \& 
Taylor 2003), 24 August 2002 and 22 August 2003. Table~\ref{tab:observations} 
lists the observing frequencies for each epoch. The observations for 
24 August 2002 and 22 August 
2003 were obtained as part of a multi-frequency polarization study of about 
three dozen BL~Lac objects. In all cases the sources were observed in a 
`snap-shot' mode with 8--10 scans of each object spread out over the time. 
The preliminary calibration, D-term calibration and polarization calibration 
were all done in AIPS using standard techniques. For more detailed calibration 
information for the 1997 data, see Gabuzda \& Chernetskii (2003) and 
Reynolds, Cawthorne \& Gabuzda (2001), and for the June 2000 data see 
Zavala \& Taylor (2003). 

Note that Gabuzda \& Chernetskii (2003) presented 
results at 22 GHz in addition to the 15, 8.4 and 5 GHz results considered 
here; we did not include the 22 GHz data in our Faraday rotation analysis 
because this substantially restricted the region in the jet where polarization 
was reliably detected at all the frequencies used. In regions where 
polarization was detected at 22, 15, 8.4 and 5 GHz, the derived rotation 
measures are consistent with the three-frequency results
presented here.

\subsection{24 August 2002 and 22 August 2003}
The instrumental polarizations (D-terms) were determined using the AIPS task 
LPCAL, solving simultaneously for the source polarization in individual VLBI 
components. The sources used for this purpose were 1308+326 (August 2002) 
and 1156+295 (August 2003).
The Electric Vector Polarization Angle (EVPA) calibration was done using 
integrated polarization observations of a bright, compact source, obtained 
with the Very Large Array (VLA), by forcing the EVPA for the total VLBI 
polarization of the source to match the EVPA for the integrated polarization 
(www.aoc.nrao.edu/~smyers/calibration/). The sources used for the EVPA 
calibration were 1749+096 (observed with the VLA on 10 August 2002) and 
2200+420 (observed with the VLA on 18 August 2002 and 21 August 2003). The sources were observed with the VLA at frequencies 5, 8.5, 22 and 43 GHz. We found the obtained values to be consistent with the (Faraday Rotation) linear $\lambda^{2}$ law and were thus able to interpolate from the graph the corresponding values for our non-standard frequencies.
We refined the calibration by checking for 
self-consistency between the various frequencies, and also with the other 
EVPA calibration results for VLBA experiments before and after ours that were 
calibrated using the same reference antenna (the EVPA corrections for a given 
reference antenna are typically stable to within $\chi\sim$5$^{\circ}$ over 
several years; see Reynolds et al. (2001), 
http://www.physics.purdue.edu/astro/MOJAVE/sourcepages). 
Table~\ref{tab:corrections} shows the final applied EVPA corrections.

\begin{figure*}
\centering
\includegraphics[width=0.7\textwidth]{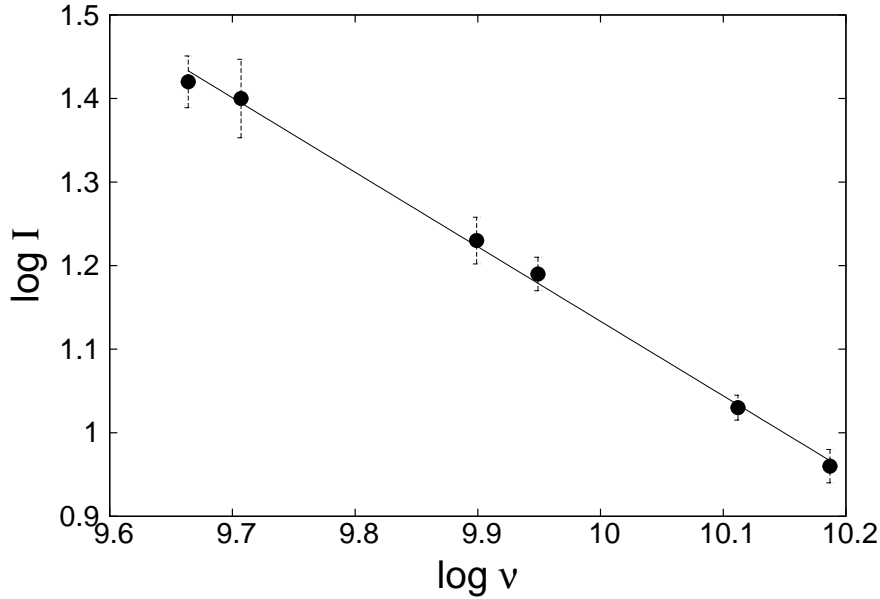}
\caption[Short caption for figure]{\label{fig:check_calibration}
Plot of $\log I_{\nu}$ (mJy) vs. $\log\nu$ (Hz) for an $3\times 3$-pixel
area in the optically thin jet of B1803+784. The observed spectrum is
fully consistent with a power-law, with spectral 
index $-0.89\pm 0.02$. There is no evidence for deviations from this
behaviour for any of the six observing frequencies.}
\end{figure*}

The August 2003 data include several frequencies that differ from the
standard 5, 8 and 15 GHz frequencies for the VLBA. Although system temperatures
and gain curves were provided for all six frequencies observed at this
epoch, there may be some concern that the accuracy of the overall flux
calibration for these slightly `non-standard' frequencies could be slightly
lower than for the standard VLBA frequencies. Accordingly, we determined
the spectra of various optically thin regions in the jet, to verify that
there was no evidence for any of the frequencies deviating from the behaviour
shown by the others. Before determining the spectra, we aligned the images
at the various frequencies using the algorithm of Croke \& Gabuzda (2008).
A typical example, corresponding to a position 3.4~mas
West and 0.3~mas North of the VLBI core, is shown in 
Fig.~\ref{fig:check_calibration}; we can see that the observed 4.6--15.3~GHz
fluxes are all consistent with a power-law within the errors, corresponding
to a ``normal'' optically thin spectral index of $-0.89\pm 0.02$. Thus, we
find no evidence for inaccuracy of the overall flux calibration at any of
our six frequencies.

\begin{table*}
\caption{Observational Parameters} \centering
\label{tab:observations}
\begin{tabular}{ll}
\hline
Epoch   & Frequency (GHz)   \\\hline
6 April 1997      & 4.990, 8.415, 15.285, 22.221${\dagger}$ \\
11 June 2000      & 8.114, 8.209, 8.369, 8.594, 12.115, 12.591, 15.165${\ddagger}$\\
24 August 2002    & 15.286, 22.234, 43.136\\
22 August 2003    & 4.612, 5.092, 7.916, 8.883, 12.939, 15.383 \\
\hline
\multicolumn{2}{l}{$^{\dagger}$Gabuzda \& Chernetskii 2003}\\
\multicolumn{2}{l}{$^{\ddagger}$Zavala \& Taylor 2003}\\
\end{tabular}
\end{table*}
\hfill

\begin{table*}
\caption{Electric Vector Position Angle Corrections (EVPAs)} \centering
\label{tab:corrections}
\begin{tabular}{lccc}
\hline
Epoch & Frequency & Reference & EVPA correction $\Delta\chi$ \\
      &   (GHz)   & Antenna   &    (deg) \\ \hline
24 August 2002 & 15.2855 & LA & $+92$  \\
               & 22.2346 & LA & $+35$  \\
               & 43.1355 & LA & $+75$  \\
22 August 2003 &  4.6120 & LA & $-66$ \\
               &  5.0920 & LA & $-66$ \\
               &  7.9160 & LA & $+103$ \\
               &  8.8830 & LA & $+92$ \\
               & 12.9390 & LA & $+81$ \\
               & 15.3830 & LA & $+109$ \\
\hline
\end{tabular}
\end{table*}

We made maps of the distribution of the total intensity $I$ and Stokes 
parameters $Q$ and $U$ at the different frequencies with matched resolutions 
corresponding to the lowest- or middle-frequency beam. The distributions of the 
polarized flux ($p = \sqrt{Q^2 + U^2}$) and polarization angle 
($\chi = \frac{1}{2}\arctan \frac{U}{Q}$) were obtained from the $Q$ and $U$ 
maps using the AIPS task COMB. We then constructed maps of the RM using a modified version of the
AIPS task RM that can construct RM maps using up to 10 frequencies,
obtained from R. Zavala (we will be happy to provide this code to anyone
who is interested), after subtracting the effect of the integrated
RM (presumed to arise in our Galaxy) from the observed polarization angles (the integrated RM value for B1803+784 was $-60$ $rad/m^2$, determined by Pushkarev 2001), so that any residual Faraday Rotation was due to only the 
thermal plasma in the vicinity of the AGN. The RMs and associated 
($\chi$) uncertainties were determined in a similar manner to that used by 
Gabuzda, Murray \& Cronin (2004), taking into account the errors in the 
angle calibration (3$^{\circ}$), as well as the error in the polarization 
angles (which were taken to be the rms deviation from the mean value of 3 x 3 
pixel (0.3 x 0.3 mas$^2$) area at the corresponding location). 

 \begin{figure*}
  \begin{minipage}[t]{7.5cm}
  \begin{center}
  \includegraphics[width=7.4cm,clip]{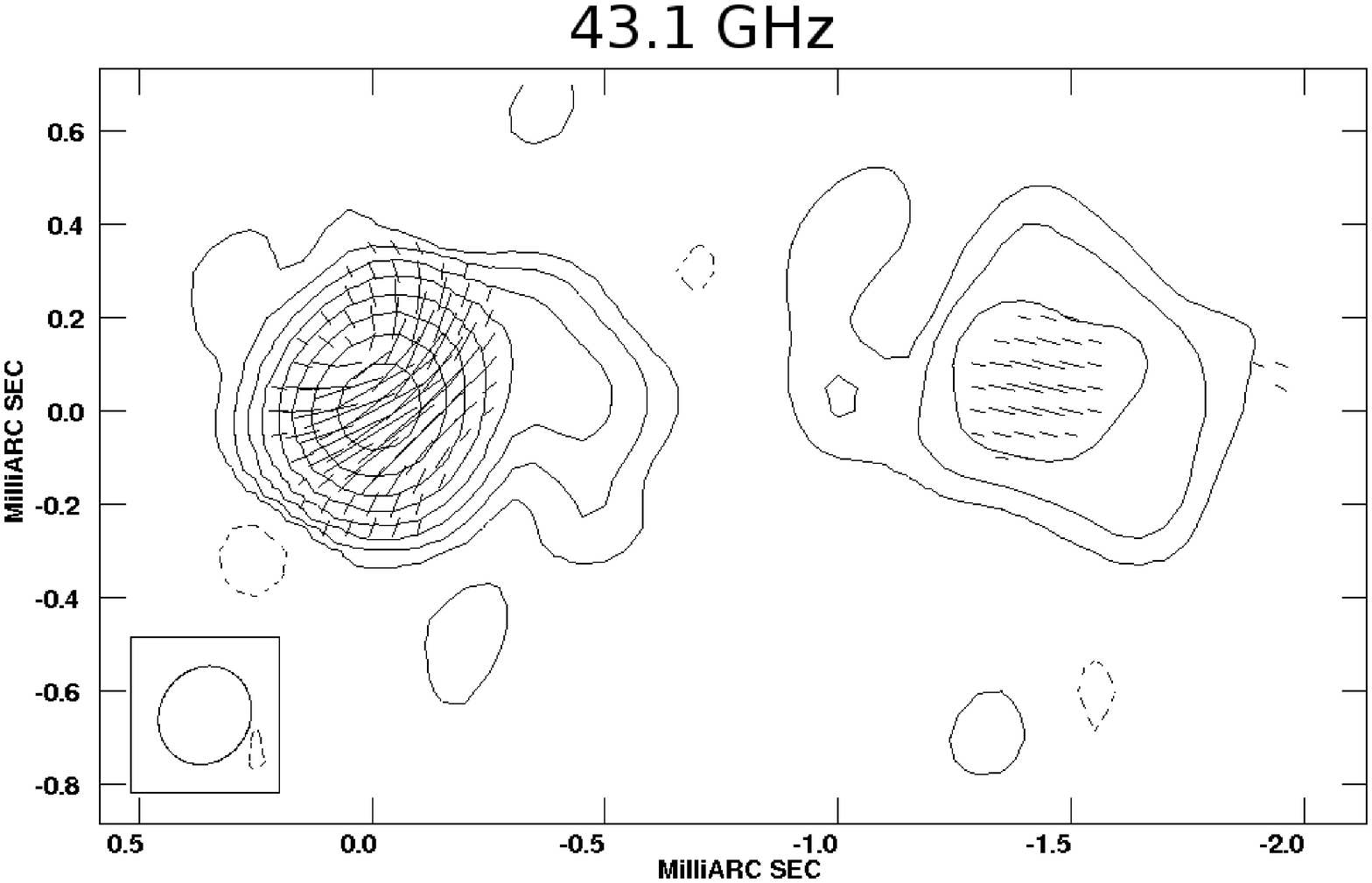}
  \end{center}
  \end{minipage}
  \begin{minipage}[t]{7.5cm}
  \begin{center}
  \includegraphics[width=7.4cm,clip]{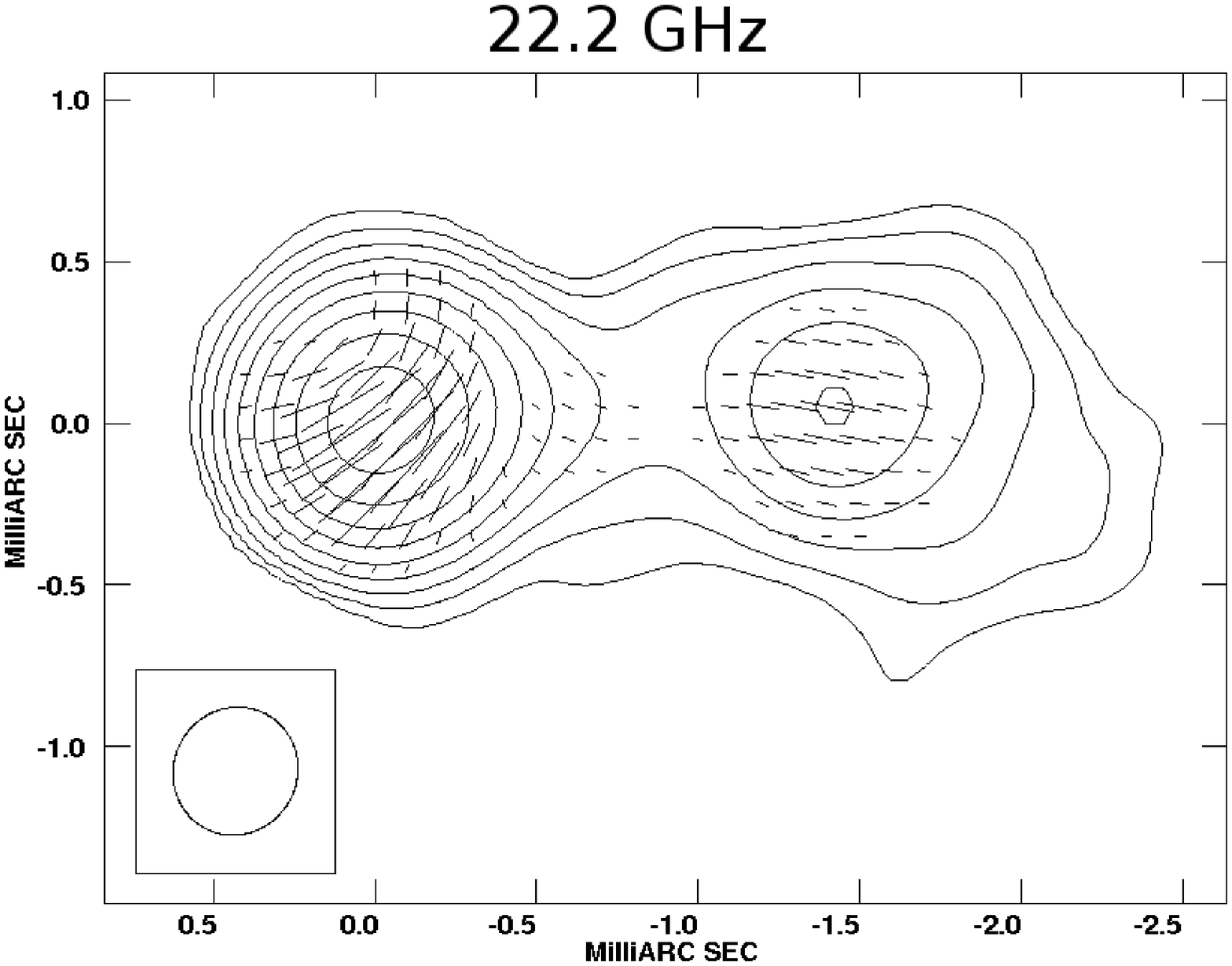}
  \end{center}
  \end{minipage}
\begin{minipage}[t]{14.0cm}
  \begin{center}
  \includegraphics[width=8.0cm,clip]{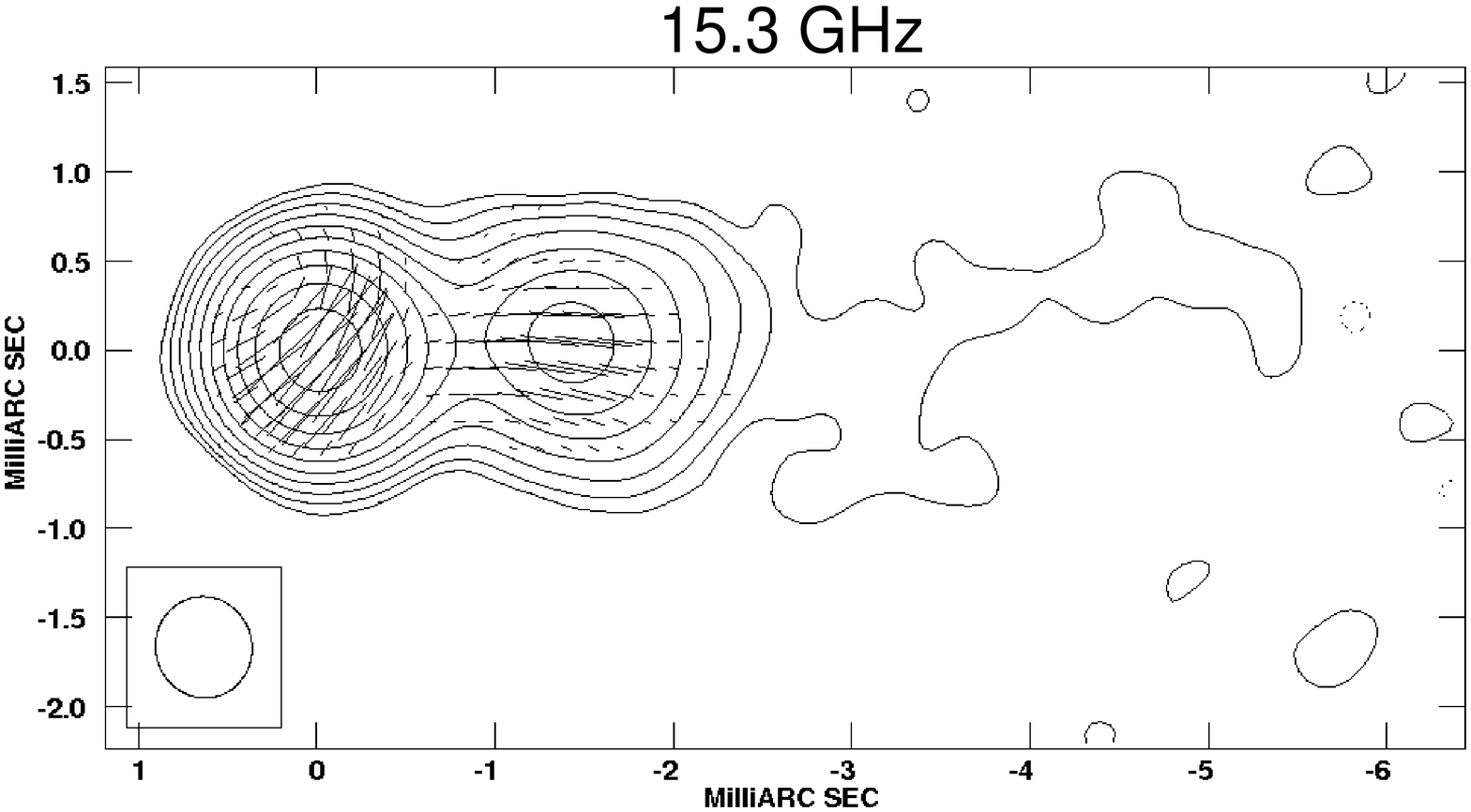}
  \end{center}
  \end{minipage}
 \caption[Short caption for figure 1]{\label{fig:slava_maps}VLBA $I$ maps with (Electric Vector) polarization sticks superimposed at 43.1, 22.2 and 15.3~GHz at epoch 24 August 2002, corrected for integrated Faraday Rotation.}
\end{figure*}

\begin{figure*}
\begin{minipage}[t]{7.5cm}
 \begin{center}
 \includegraphics[width=7.5cm,clip]{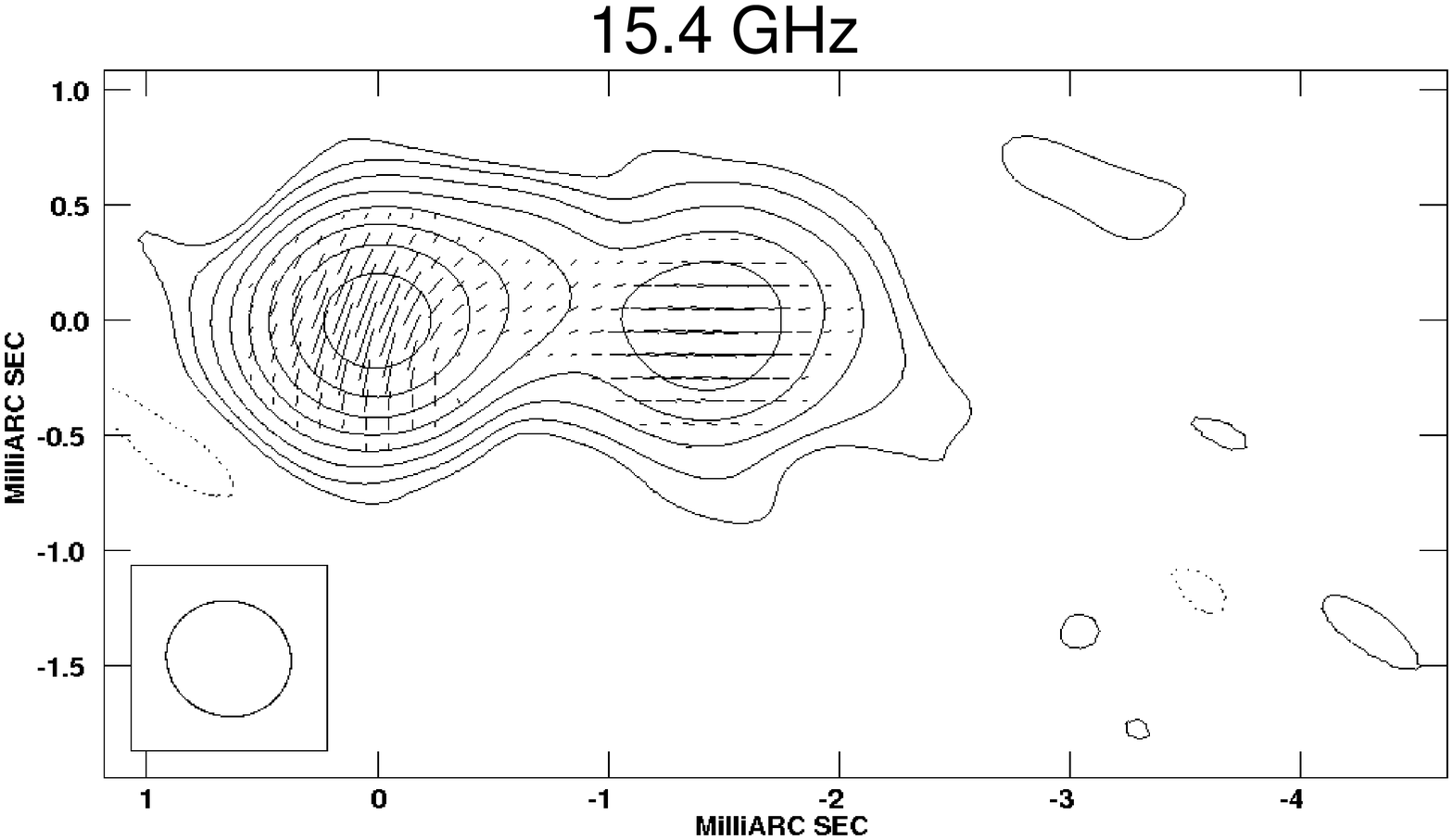}
 \end{center}
\end{minipage}
\begin{minipage}[t]{7.5cm}
 \begin{center}
 \includegraphics[width=7.5cm,clip]{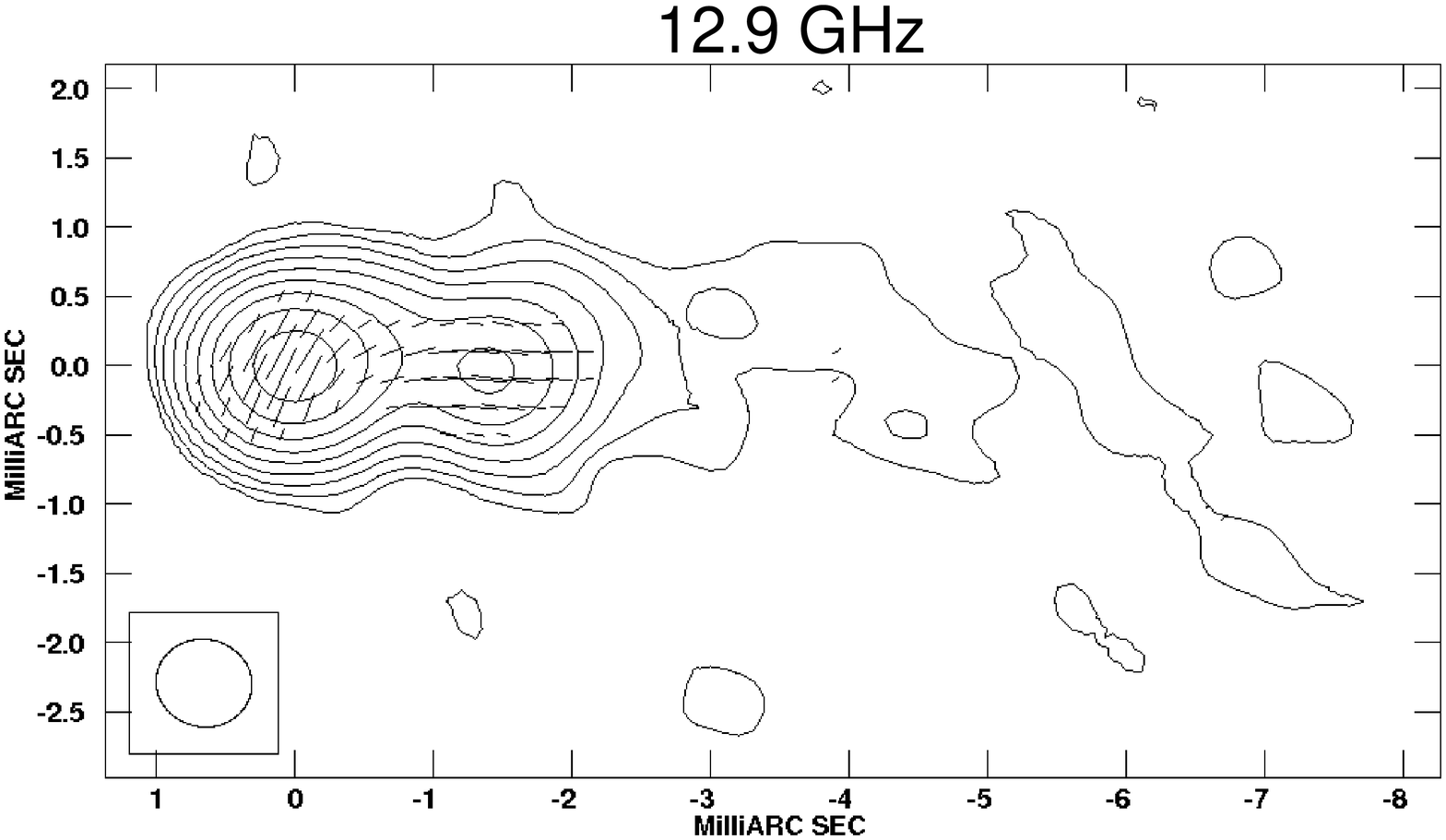}
 \end{center}
\end{minipage}
\begin{minipage}[t]{7.5cm}
 \begin{center}
 \includegraphics[width=7.5cm,clip]{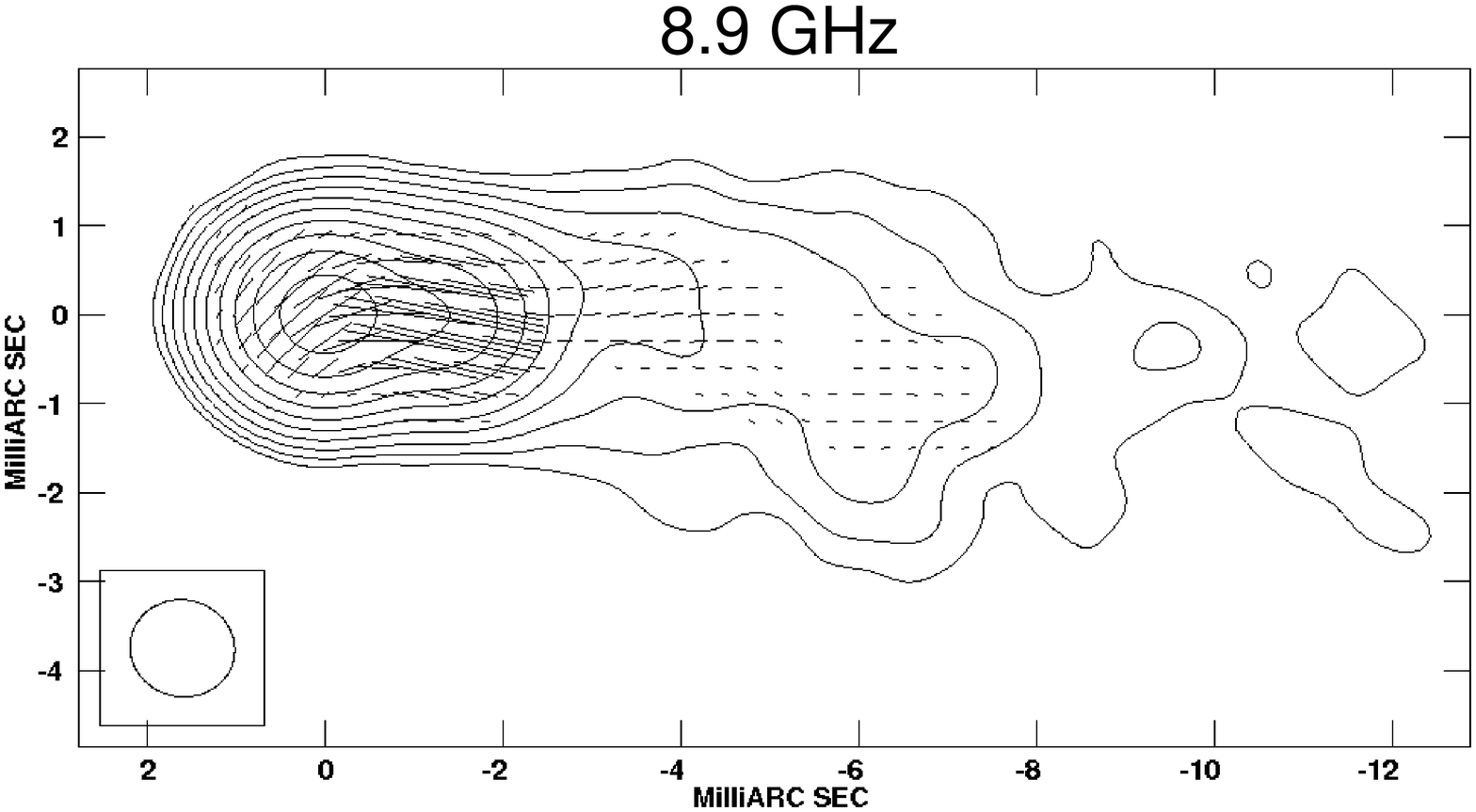}
 \end{center}
\end{minipage}
\begin{minipage}[t]{7.5cm}
 \begin{center}
 \includegraphics[width=7.5cm,clip]{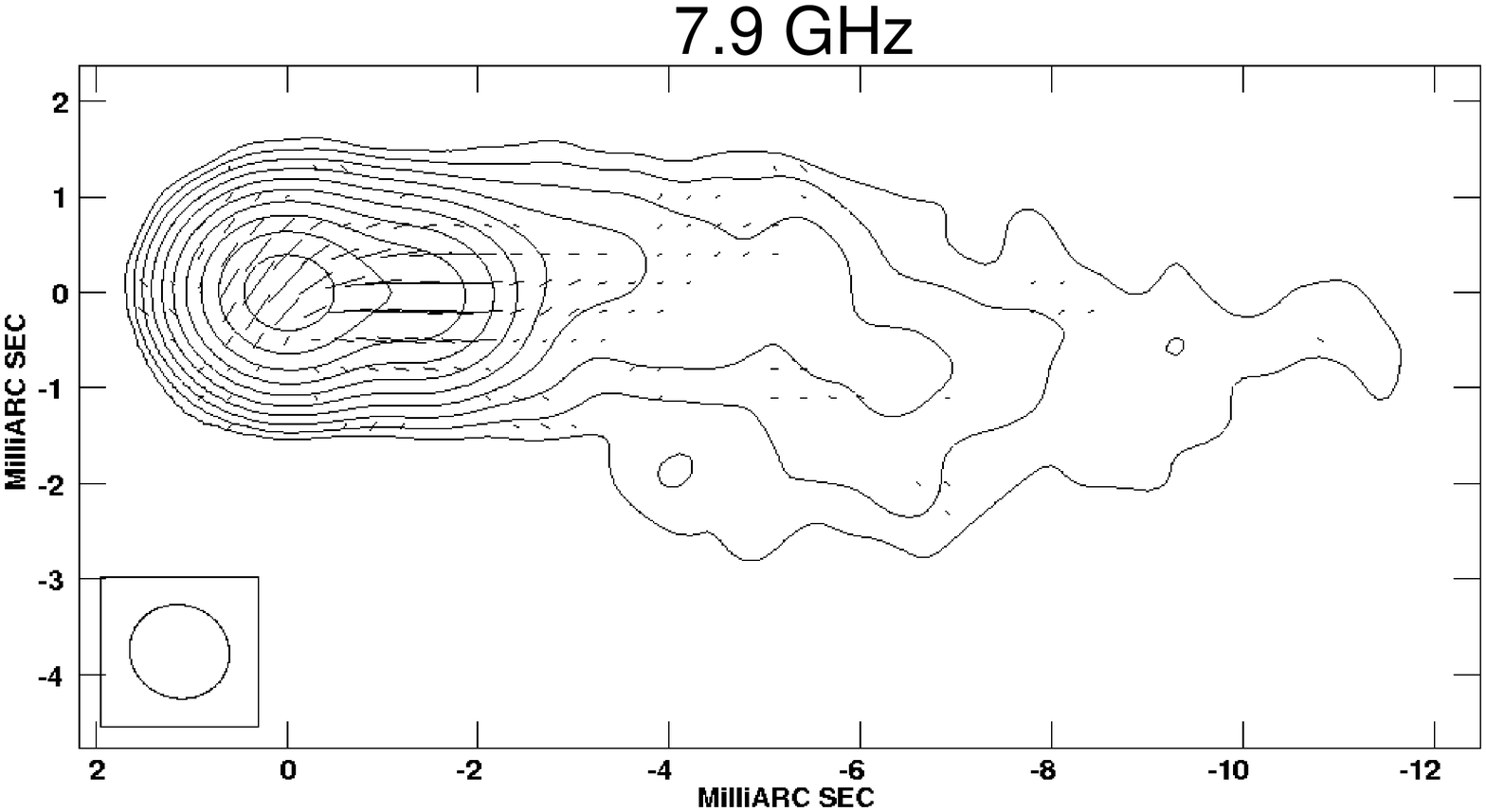}
 \end{center}
\end{minipage}
\begin{minipage}[t]{13.7cm}
 \begin{center}
 \includegraphics[width=13.7cm,clip]{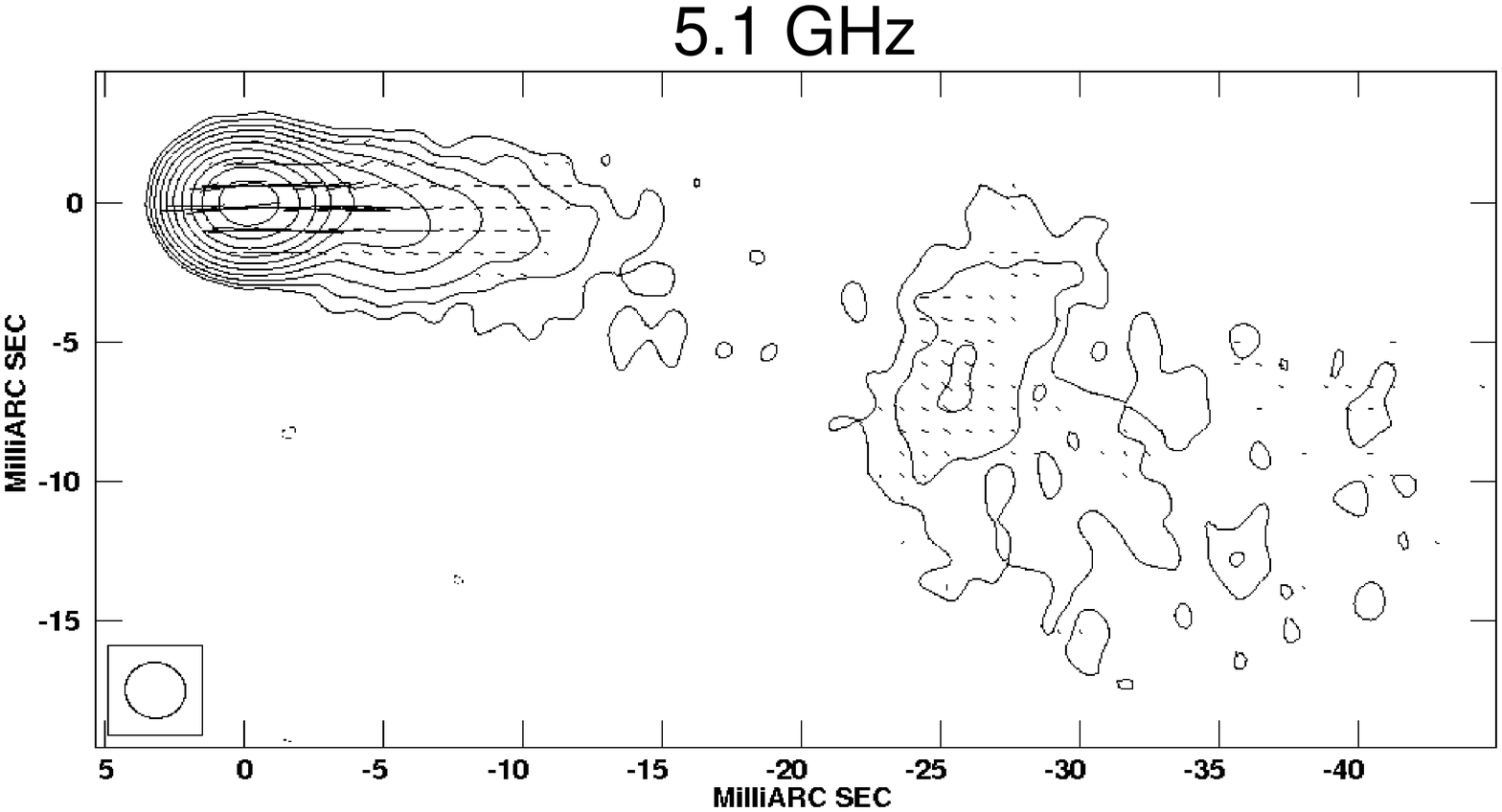}
 \end{center}
\end{minipage}
 \begin{minipage}[t]{13.7cm}
 \begin{center}
 \includegraphics[width=13.7cm,clip]{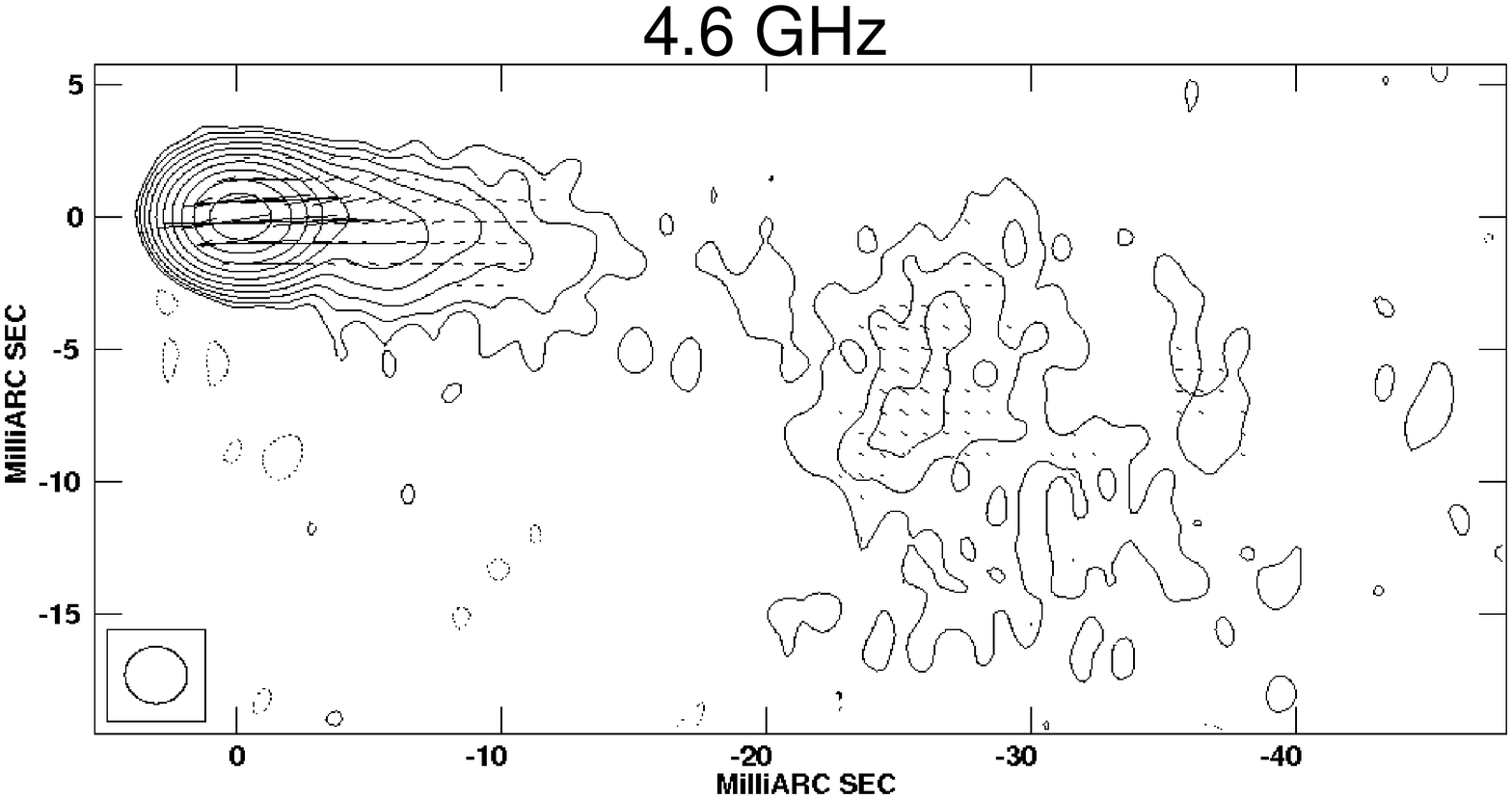}
 \end{center}
 \end{minipage}
\caption[Short caption for figure 2]{\label{fig:my_pol_maps}VLBA $I$ maps with 
(electric vector) polarization sticks superimposed at all six frequencies (See 
 Table~\ref{tab:observations}) at epoch 22 August 2003, corrected for 
 integrated Faraday Rotation.}
\end{figure*}

\begin{figure*}
\centering
\includegraphics[width=1.0\textwidth]{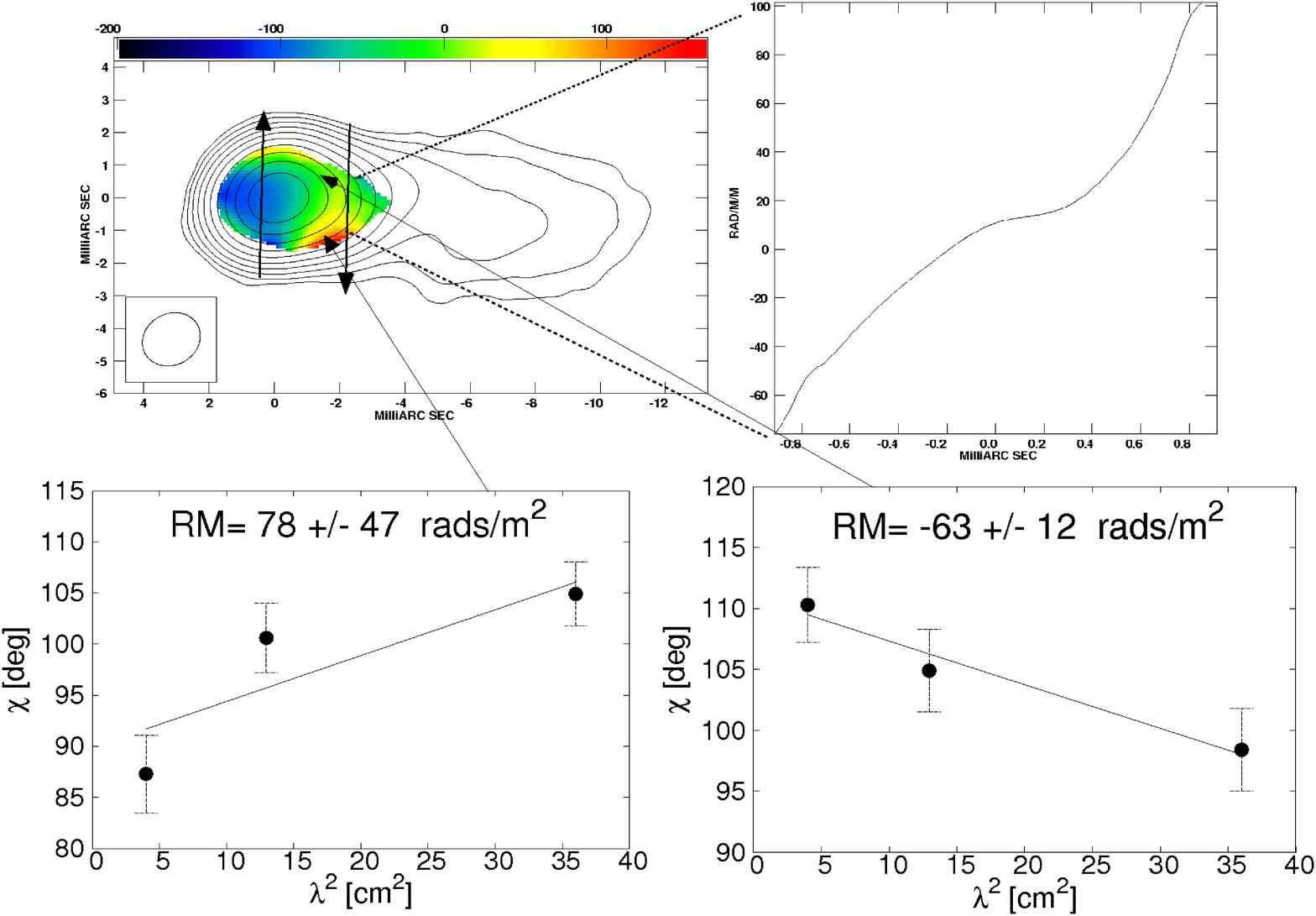}
\caption[Short caption for figure]{\label{fig:apr1997jet}RM map of B1803+784 observed on 6 April 1997. The range of RM values shown in the color bar is in units of rad/m$^2$. The $I$ contours are those for 8.4 GHz. The peak flux (Jy/beam) is 1.9~Jy/beam and the lowest contour is 2.0~mJy/beam. The accompanying panels show a slice of the RM distribution across the jet, and plots of the polarization angle $\chi$ vs. $\lambda^2$ for the indicated 3 x 3 pixel regions near the top and bottom of the jet. The errors shown are 1 $\sigma$.}
\end{figure*}

\begin{figure*}
\centering
\includegraphics[width=1.0\textwidth]{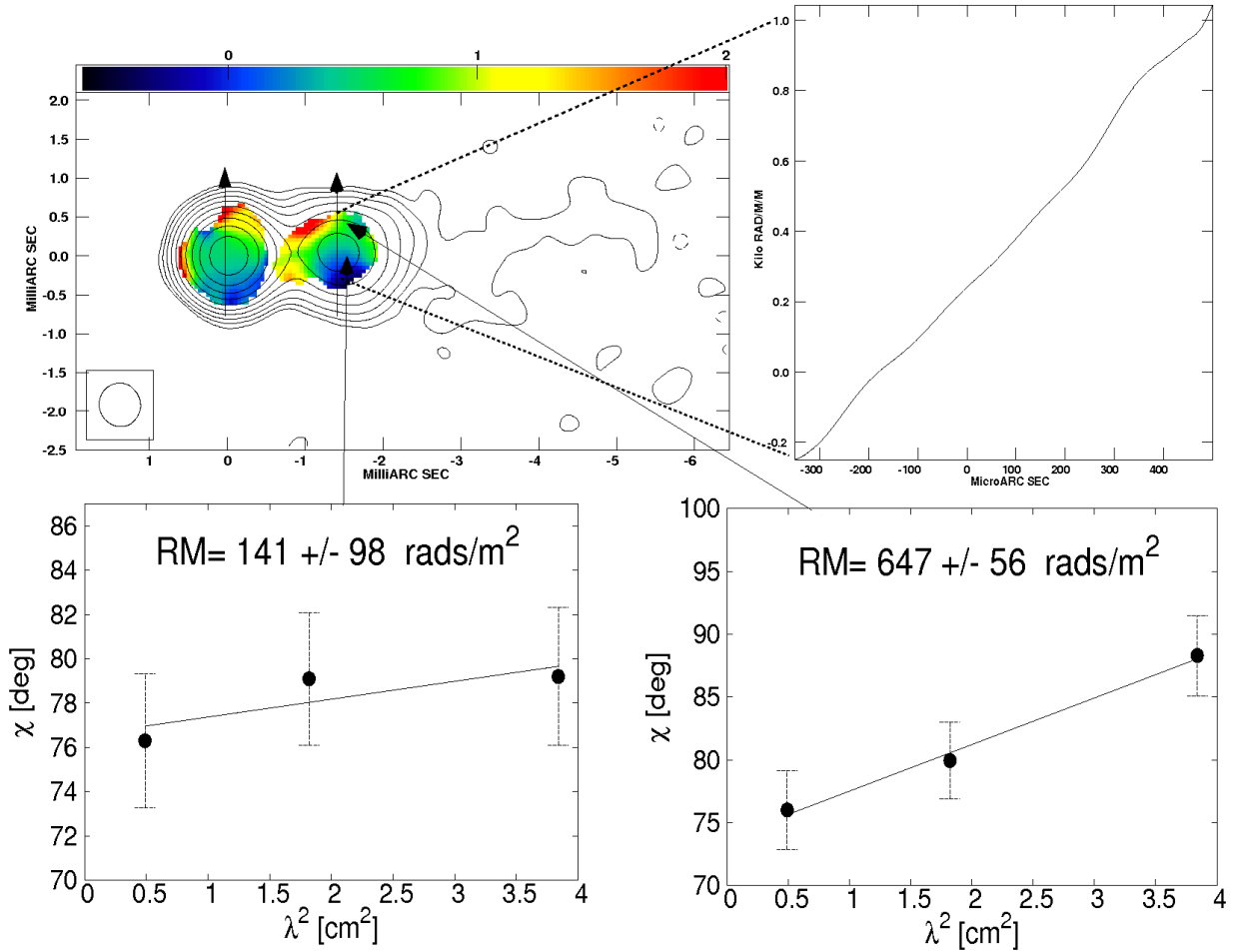}
\caption[Short caption for figure]{\label{fig:aug2002jet}RM map of B1803+784 
observed on 24 August 2002. The range of RM values shown in the color bar is in units of kilorad/m$^2$. The $I$ contours are those for 15.3 GHz. The 
peak flux is 1.7~Jy/beam and the lowest contour is 2.0~mJy/beam. 
The accompanying panels show a slice of the RM distribution across the jet, 
and plots of polarization angle $\chi$ vs. $\lambda^2$ for the indicated 
3 x 3 pixel regions near the top and bottom of the jet. The errors shown 
are 1 $\sigma$.}
\end{figure*}

\begin{figure*}
\centering
\includegraphics[width=1.0\textwidth]{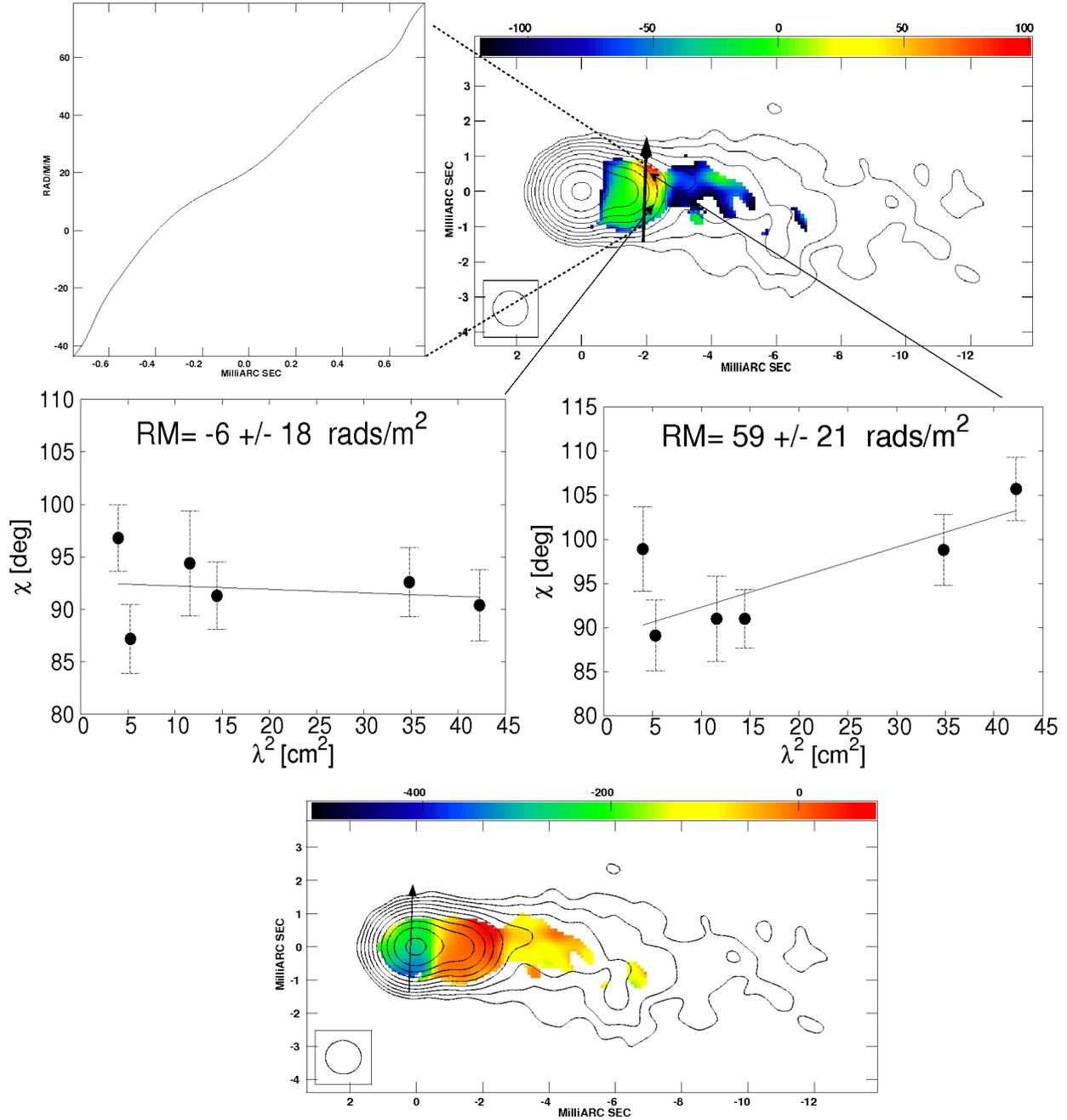}
\caption[Short caption for figure]{\label{fig:aug2003jet}RM maps of B1803+784
observed on 22 August 2003; the two maps correspond to the same RM
distribution, but are shown separately with different RM ranges to highlight 
the RM distributions in the jet (top map) and core (bottom map). The range of RM values shown in the color bars is in units of rad/m$^2$. The $I$ contours are those at 7.9 GHz. The peak flux is 1.2~Jy/beam and the 
lowest contour is 2.0~mJy/beam. The accompanying panels show a slice of the 
RM distribution across the jet, and polarization angle $\chi$ vs. 
$\lambda^2$ plots for the indicated 3 x 3 pixel regions near the 
top and bottom of the jet. The errors shown are 1 $\sigma$.}
\end{figure*}

\begin{table*}
\caption{Map Parameters of Figure~\ref{fig:slava_maps} and Figure~\ref{fig:my_pol_maps}} \centering
\label{tab:maps}
\begin{tabular}{lll}
\hline
Epoch  & Peak (Jy/beam) ${\dagger}$  & Bottom contour (mJy/beam) ${\dagger}$   \\\hline
24 August 2002 & 1.7, 1.6, 0.9  &  0.2, 0.4, 4.5 \\
22 August 2003 & 1.3, 1.2, 1.3, 1.3, 1.3, 1.2   &  1.6, 1.5, 1.6, 1.6, 3.3, 5.8 \\
\hline
\multicolumn{3}{l}{$^{\dagger}$ Listed in order of ascending frequency; see Table~2.}\\
\end{tabular}
\end{table*}
\hfill

\begin{figure*}
\begin{minipage}[t]{7.5cm}
 \begin{center}
 \includegraphics[width=7.5cm,clip]{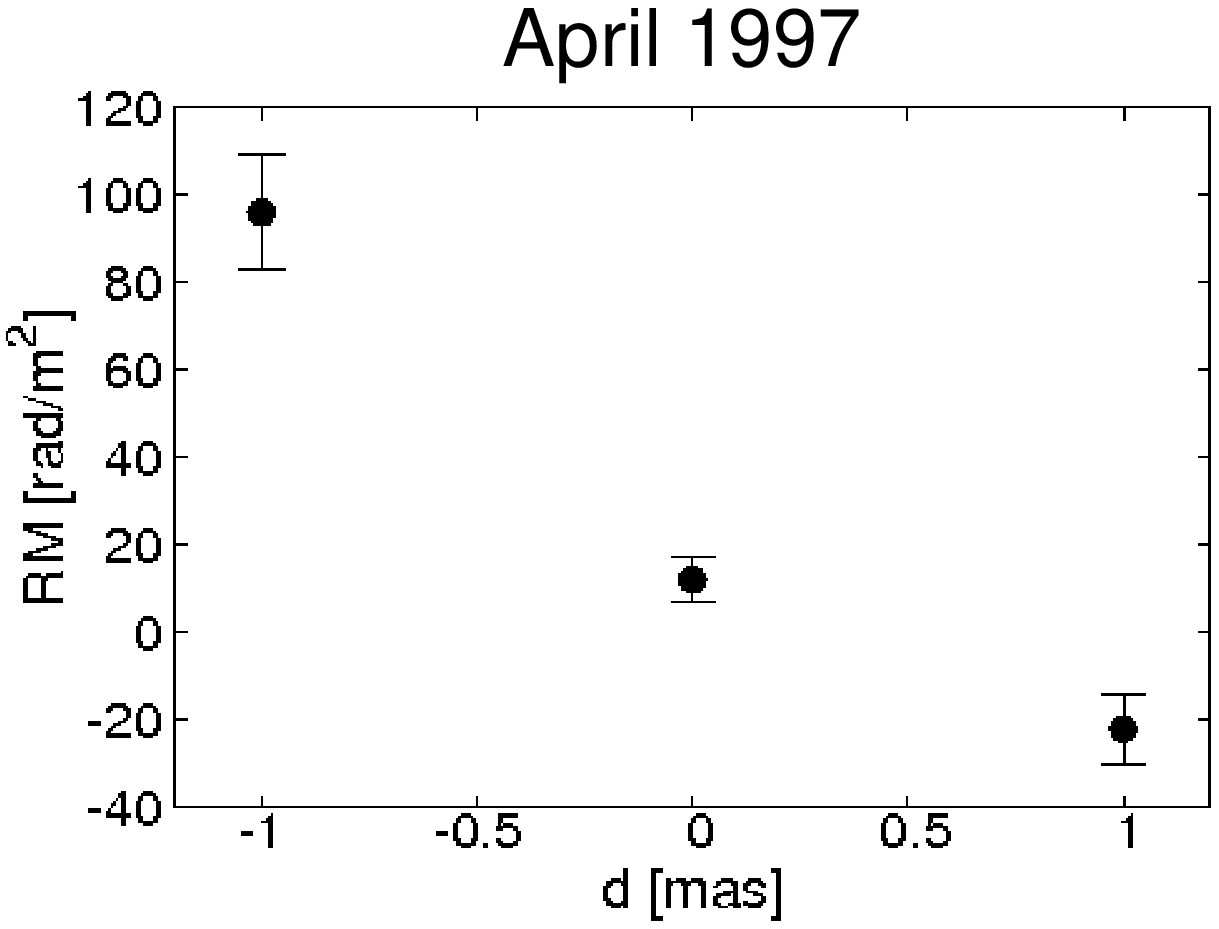}
 \end{center}
\end{minipage}
\begin{minipage}[t]{7.5cm}
 \begin{center}
 \includegraphics[width=7.5cm,clip]{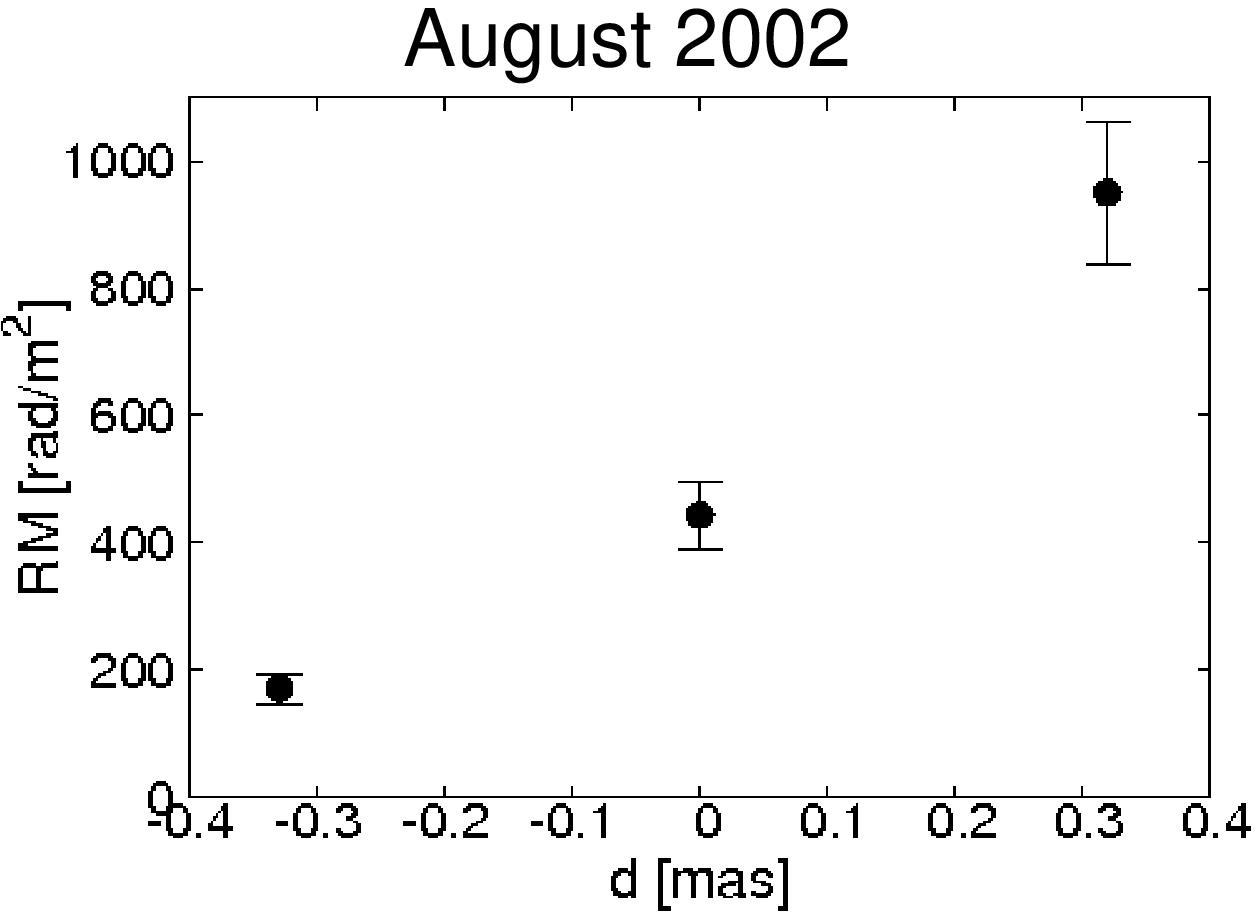}
 \end{center}
\end{minipage}
 \begin{minipage}[t]{7.5cm}
 \begin{center}
 \includegraphics[width=7.5cm,clip]{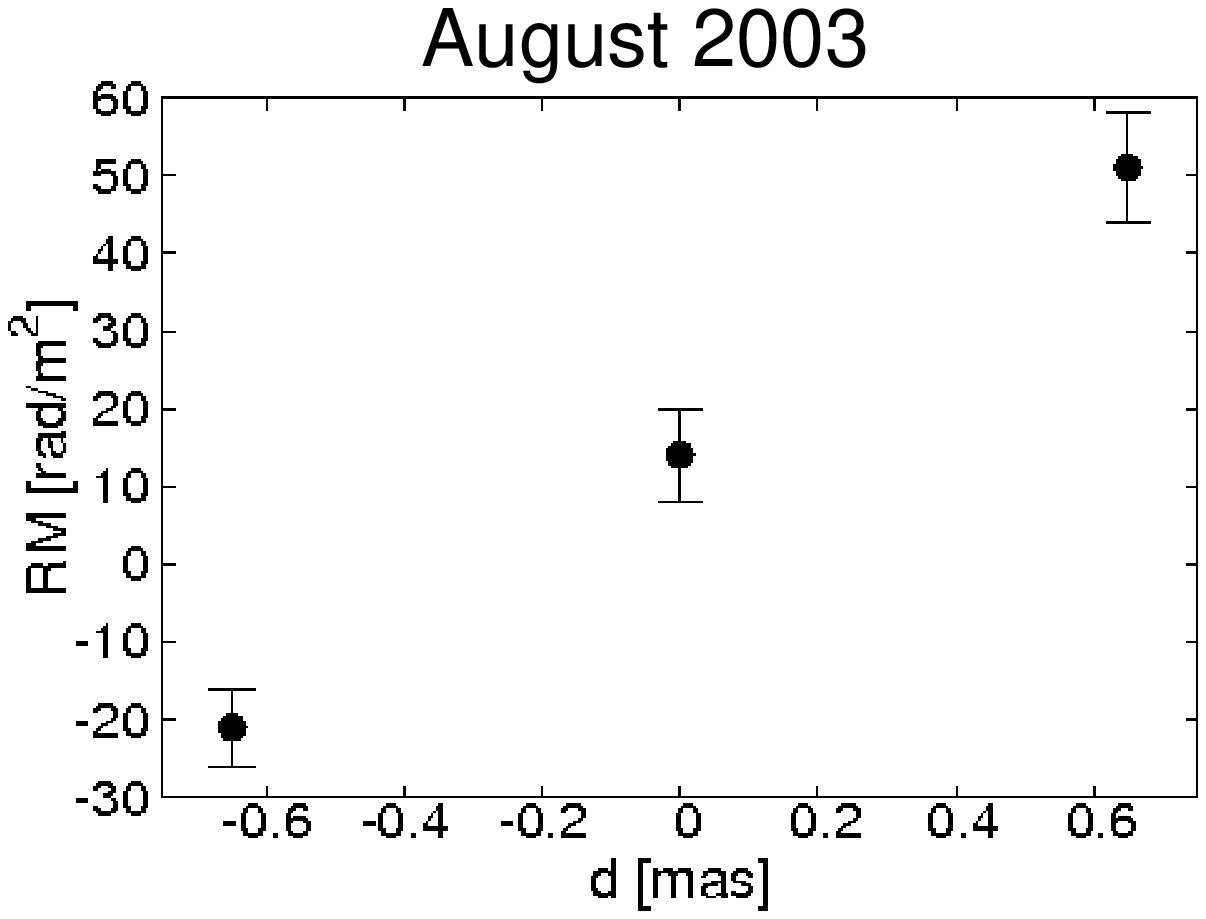}
 \end{center}
 \end{minipage}
 \caption[Short caption for figure 2]{\label{fig:trans_dist} Plots of observed 
RM as a function of transverse distance {\emph d} from a reference point 
{\bf P} near the central jet spine, for cuts from North to South across the
jet and core region of B1803+784 for the three epochs. For April 1997: 
{\bf P} = (-1.7 mas,-0.1 mas); for August 2002: {\bf P} = (-1.4 mas,-0.13 mas) 
and for August 2003: {\bf P} = (-1.9 mas,-0.05 mas). The distance $d$ is
positive to the North of {\bf P} and negative to the South of {\bf P}. 
All errors are 1 $\sigma$.}
\end{figure*}

\section{Results}
The total intensity ($I$), linear polarization ($P$) and RM maps for epoch 
11 June 2000 are presented by Zavala \& Taylor (2003) and $I$ and $P$ 
maps for 6 April 1997 by Gabuzda \& Chernetskii (2003); we present the RM 
map for the latter epoch below. An angular distance of 1~mas corresponds to 
a linear distance of 7.06~pc at the redshift (0.68) of B1803+784 (http://www.physics.purdue.edu/MOJAVE/sourcepages/1803+784.shtml).

Figs.\ \ref{fig:slava_maps} and \ref{fig:my_pol_maps} show the observed VLBI 
total intensity and linear polarization structures for August 2002 and August 
2003. The peaks and bottom contours of these maps are given in 
Table~\ref{tab:maps}.
The images in Figs.\ \ref{fig:apr1997jet}, \ref{fig:aug2002jet} and 
\ref{fig:aug2003jet} show the VLBI total intensity contours for 6 April 1997, 
24 August 2002 and 22 August 2003 with their parsec-scale RM distributions 
superimposed. In all the maps presented here, the contours increase in 
steps of a factor of two, and the beam is shown in the lower left corner 
of the image.

The polarization maps for all four epochs are consistent with a predominantly 
transverse magnetic-field structure. Transverse RM gradients are visible 
across the VLBI jet of B1803+784 in each of the three epochs shown in
Figs.\ref{fig:apr1997jet}--\ref{fig:aug2003jet}, on scales of 
$\sim 2-3$ mas; we are also tentatively able to follow the RM gradient
further from the core in Fig.~\ref{fig:aug2003jet}. This supports the 
hypothesis that this jet has a helical 
magnetic field, consistent with the observed transverse magnetic-field 
structure. The arrows show the directions of the RM gradients in the 
corresponding regions; in other words, the direction in which the value of
the RM increases (from more negative to less negative, negative to positive,
or less positive to more positive, as the case may be). 
The accompanying panels show plots of 
polarization angle ($\chi$) vs. wavelength squared ($\lambda^2$) for the 
indicated regions, as well as slices through the RM distributions between the 
indicated points. The uncertainties in the polarization angles are also shown 
in the plots. 

Fig.~\ref{fig:trans_dist} shows plots of the observed RM as a function of the 
transverse distance from the central spine of the jet. The errors in the RMs 
in these plots were estimated in the same way as for the $\chi$ values, by 
calculating the mean RM within the corresponding $3\times 3$~pixel area in the 
map and assigning the rms deviation about this mean as the RM error. We 
consider this approach to be most reasonable, since possible EVPA calibration 
errors will affect the fitted RM values and the uncertainties derived from the 
fits, but not the inferred RM gradients, as is shown in the Appendix. Both 
these plots and the RM slices in 
Figs.~\ref{fig:apr1997jet} -- \ref{fig:aug2003jet} clearly demonstrate the 
systematic, monotonic nature of the observed RM gradients.

The RM map of Zavala \& Taylor (2003) for June 2000 also shows a clear 
transverse gradient, with a negative RM along the Northern edge of the jet 
and less negative or positive RM along the Southern edge, similar to the RM 
map observed in August 2002 (Fig.~\ref{fig:aug2002jet}). An important feature 
that emerges in a comparison of the various RM maps is the reversal in the 
direction of the RM gradient in the jet of B1803+784 between June 2000 and 
August 2002. The RM values increase toward the Southern edge of the jet 
in the April 1997 and June 2000 (Zavala \& Taylor 2003) RM maps, but toward the Northern edge of 
the jet in the August 2002 and August 2003 RM maps. 

\begin{figure*}
\centering
\includegraphics[width=0.5\textwidth]{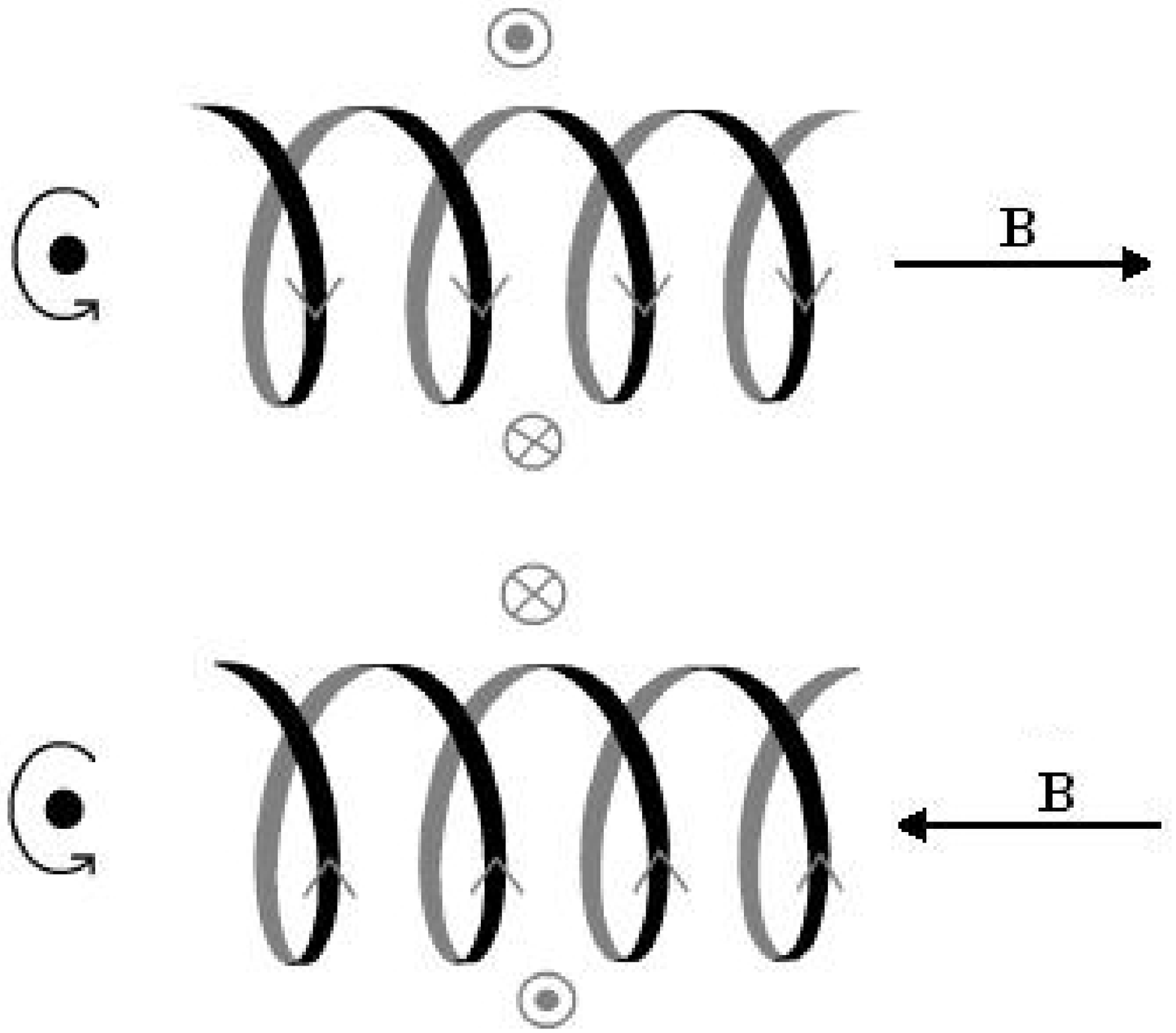}
\caption[Short caption for figure]{\label{fig:north_south_rev}Illustration of 
how a ``flip'' in the RM gradient can be caused by a change in the pole of 
the black hole facing the earth. The arrows mark the direction of the B-field, 
and are marked on the nearer side of the helical field to the observer.}
\end{figure*}

\begin{figure*}
\centering
\includegraphics[width=1.0\textwidth]{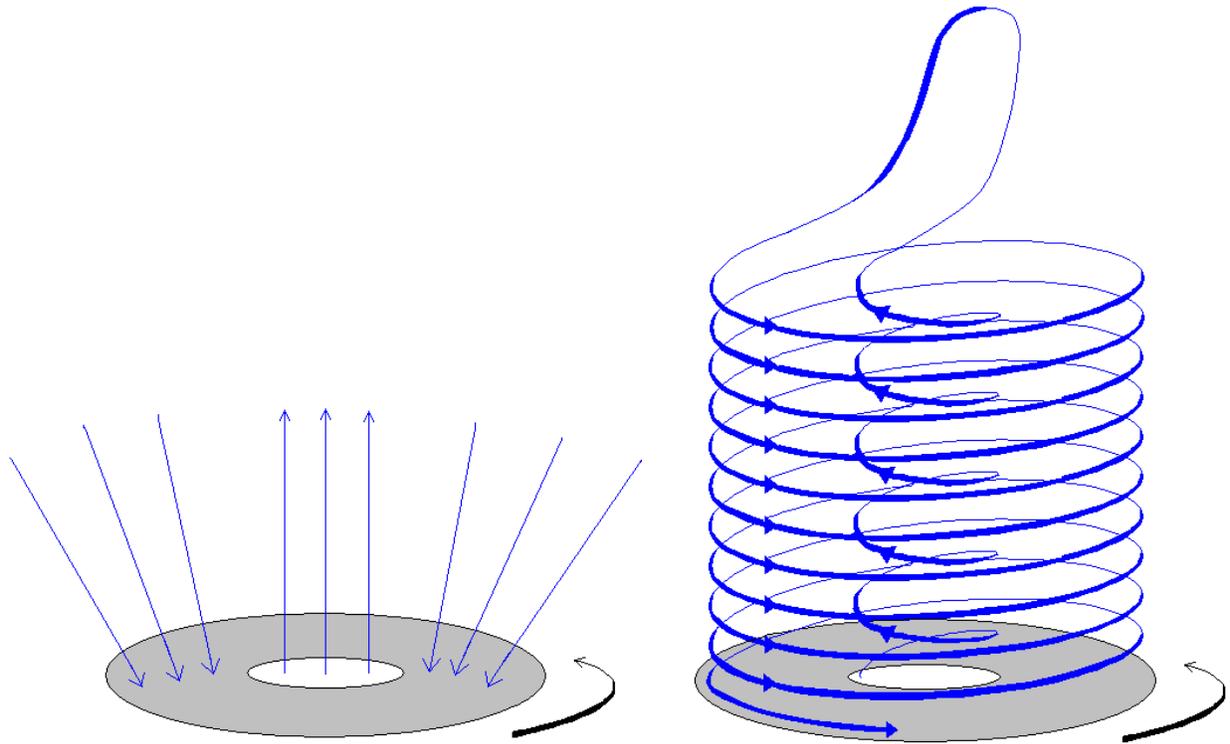}
\caption[Short caption for figure]{\label{fig:nested_hlx}Illustration showing 
how the magnetic field lines from a magetic tower model (with the magnetic 
field lines going in the direction of outflow,finally looping back down to 
the accretion disk) can get wound up as an ``inner'' and ``outer'' helix, as 
a result of the differential disk rotation.The first figure shows the 
direction of the magnetic field lines in this model, whereas the second 
follows the path of one of these magnetic field lines that gets wound up in 
a ``nested helical field'' with the ``inner'' helix less tightly wound than 
the ``outer'' helix.}
\end{figure*}

\begin{figure*}
\centering
\includegraphics[width=1.0\textwidth]{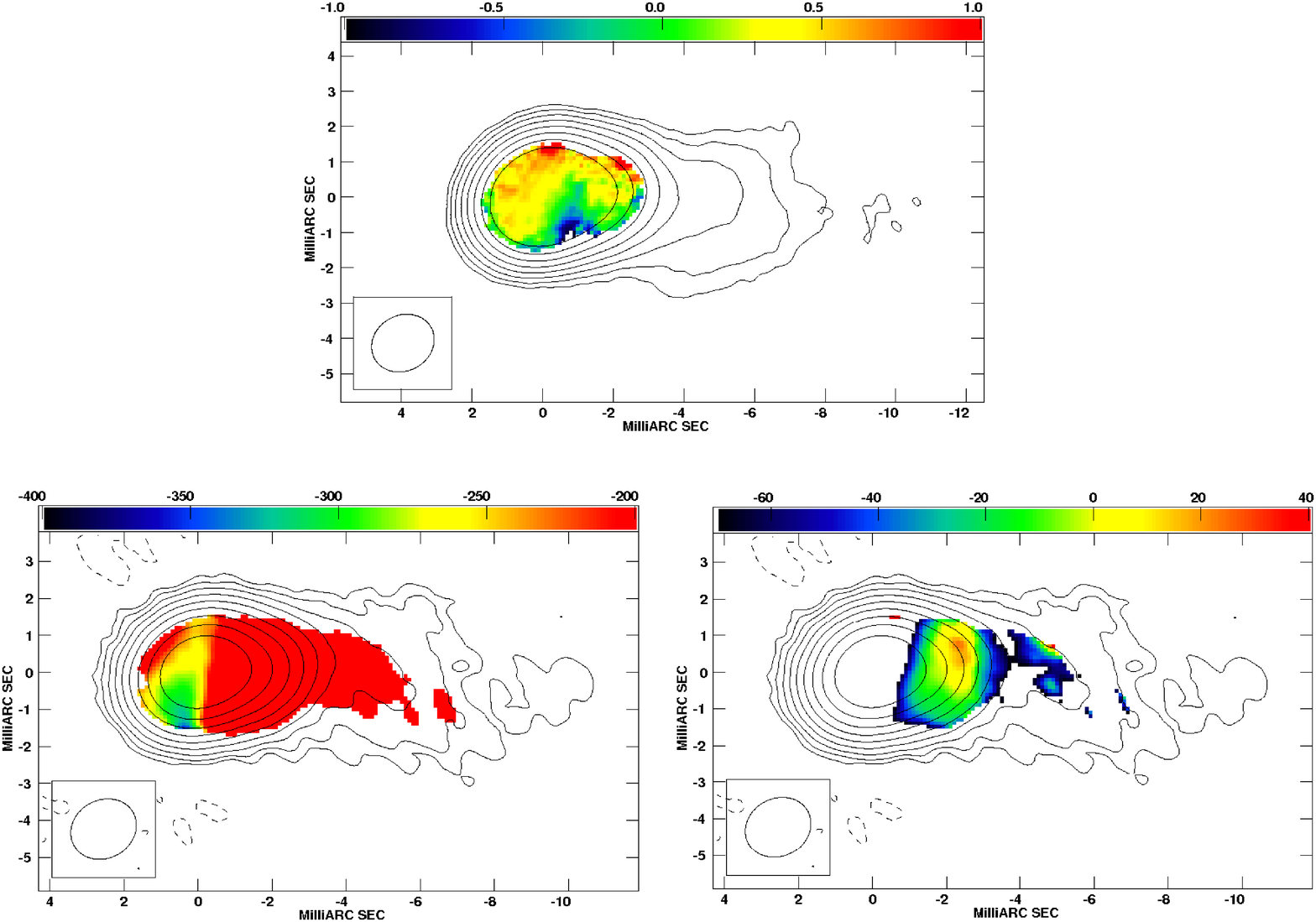}
\caption[Short caption for figure]{\label{fig:same_beam}RM maps of B1803+784
observed on August 2002 (top) and August 2003 (core region shown on bottom 
left and jet on bottom right), convolved with the same beam as was used for the April 1997 image in Fig.~\ref{fig:apr1997jet}. The range of RM values shown in the color bars for the RM maps is in units of kilorad/m$^2$ (top: August 2002) and in rad/m$^2$ (bottom left and bottom right: August 2003). The peak fluxes are 
1.9~Jy/beam for August 2002 and 1.4~Jy/beam for August 2003; the lowest 
contours are 2.0~mJy/beam for both epochs. The transverse gradients identified based
on the images with their intrinsic beams are still visible in each of
these matched-resolution maps.}
\end{figure*}

We also observe gradients in the core region for all 3 of our epochs, shown 
by the arrows in the core region in Figs.~\ref{fig:apr1997jet}, 
\ref{fig:aug2002jet} and \ref{fig:aug2003jet}, all 
of which are in the same direction (i.e., the ``flip'' of the gradient 
occurred only in the jet region). There are hints of a RM gradient with the 
same direction in the core in the map presented by Zavala \& Taylor (2003), 
but the range for the colour scale of that map makes it difficult to be sure 
of this. Note that the observed ``core'' emission is actually a mixture of 
emission from the core and inner jet, so it is not surprising that transverse 
RM gradients should be detected in the core region, since the associated 
helical field is present on a wide range of scales, down to the accretion 
disk region.

\section{Discussion}

\subsection{Possible Origins of the RM-gradient Reversal}

The origin for the observed ``flip'' in the direction of 
the transverse RM gradient in the jet of B1803+784 between June 2000 and 
August 2002 is not clear. No radical change in the intensity or polarization 
structure of the VLBI jet accompanied the observed RM-gradient flip. We
qualitatively consider here several possible scenarios that could potentially
give rise to this behaviour. Although it is not possible at this stage
to distinguish between these scenarios, our goal here is simply to
point out possible explanations, which it may be possible to test with 
further observations or numerical simulations. 

In this discussion, we will suppose that the RM gradient is 
associated with a helical jet {\bf B} field that is due to the rotation of 
the central black hole.  In this case, the direction of the observed 
transverse RM gradient is determined by the direction of rotation of the 
black hole and the ``polarity'' of the poloidal field component, i.e, North 
or South/outward or inward relative to the black hole. 

{\bf Reversal of the ``pole'' of the black hole facing the Earth.} 
It is obviously not reasonable to suppose that the observed reversal in the RM 
gradient is due to a change in the direction of rotation of the central 
black hole. However, one way to retain a transverse RM gradient in a helical 
magnetic field model but reverse the direction of this gradient, is if the 
direction of rotation of the central black hole (i.e.~the direction in which 
the field threading the accretion disc is ``wound up'') remains constant, 
but the ``pole'' of the black hole facing the Earth changes from North to 
South, or vice versa (Fig.~\ref{fig:north_south_rev}). 
To our knowledge, it is currently not known whether such polarity reversals 
are possible for AGN jets, or whether they could occur on a time scale as 
short as a few years.
Numerical simulations 
of the generation and propagation of helical magnetic fields for various
initial seed-field configurations could potentially answer these questions.
 
{\bf Torsional oscillations of the jet.} 
Another way in which the direction of the azimuthal component of
a helical {\bf B} field could change is as a result of torsional oscillations 
of the jet, such as those that have been proposed by Bisnovatyi-Kogan (2007). 
These torsional oscillations, which may stabilize the jets, essentially 
correspond to a change in the direction of rotation of the jet, and could 
cause reversals of the azimuthal {\bf B} field component from time to time,
probably periodically or quasi-periodically. Of course, we have
only observed a single reversal in the direction of the RM gradient,
and so have no information at this stage about whether this phenomena
is periodic. Therefore, this model remains a possible explanation for
the observed RM-gradient reversal, but we currently have no firm evidence in
support of this model. If future multi-frequency
VLBA polarization observations show the presence of periodic reversals of
the direction of a transverse RM gradient across the jet, this will provide
support for this type of model.  

{\bf A ``Nested-helix'' magnetic-field structure.}
Another intriguing possibility is related to magnetic-tower-type scenarios
for the launching of jets (Lynden-Bell 1996). In this picture, magnetic field 
lines anchored in the inner accretion disc are carried outward along with
the jet flow and return in outer layers of the jet, closing through the 
outer part of the accretion disc. Combined with the effect of the 
differential rotation of
the accretion disc, this gives rise to a ``nested helical field''
structure, with an ``inner'' helical field displaying a helicity corresponding
to the direction of rotation of the accretion disc and the magnetic 
polarity of the jet base and an ``outer'' helical field displaying the
opposite helicity (Fig.~\ref{fig:nested_hlx}). The pitch angles of the inner 
and outer helical fields may be different, most likely with the inner field 
having a lower pitch angle (being more tightly wound). The net RM gradient 
we observe will include contributions from
both of these fields, corresponding to the volumes of both of these regions 
that lie along the line of sight between the emission region and the 
observer.  The
directions of the RM gradients associated with each region will be opposite,
and the net observed RM gradient will be determined by whether the inner or 
outer region makes the larger contribution to the net observed RM. 
In turn, this will be determined by the distribution of the electron density 
and magnetic field along the jet and with distance from the jet axis. 
The direction of the observed RM
gradient could ``flip'' if, for example, the outer region of helical field 
usually dominates the observed RM distribution, but the inner region of 
helical field temporarily became dominant due to an increase in the electron 
density or magnetic-field strength in the inner region of helical field.
Such variations in electron density and/or 
magnetic-field strength could come about due to variations in activity of
the central engine and the ejection of jet material. In this case, the
resulting reversals in the direction of the net observed RM gradient would
most likely not occur periodically, and may follow activity of the VLBI 
core.  If future observations provide evidence for a connection between 
epochs when the direction of the transverse RM gradient reverses and 
periods of enhanced core activity, this will provide support for this type 
of scenario. 

{\bf Changes in the distribution of thermal electrons surrounding the
jet.} Finally, we should consider whether the observed change in the
direction of the transverse RM gradient could be due purely to changes
in the distribution of thermal gas in the vicinity of the jet, unrelated
to the presence of a helical jet magnetic field.
Could the transverse gradient in the RM be due, for example,
to a gradient in the thermal electron density across the jet, with 
the change in the direction of the gradient corresponding to a change in 
the side of the jet where the electron density was highest? 
In this scenario, the changes in the observed RM distribution would 
essentially be due to ``patchiness'' in a moving intervening Faraday 
screen,  and so would probably occur fairly randomly with time. 

We find this explanation unlikely for several reasons. The first is that
our observations and the observations of Zavala \& Taylor (2003)
consistently show the presence of transverse RM gradients over roughly
six years, whereas we would expect the RM pattern produced by a moving
``patchy'' screen not to be as consistent. Further, in our 1997 RM map
(Fig.~\ref{fig:apr1997jet}), the RM on the Northern side of the
jet is negative, whereas the RM on the Southern side of the jet is 
positive: a change of the sign of the RM from one side of the jet to
the other is difficult to explain as the result of a patchy thermal-electron
distribution, but is quite natural if the observed RM gradient is due to
a helical jet magnetic field. With regard to the observed changes in the
RM distribution, a comparison of our 1997 and 2003 RM maps (Figs.~\ref{fig:apr1997jet} and \ref{fig:aug2003jet}) 
shows that the RM on the Northern side of the jet is negative in 1997,
but positive in 2003.  Here, also, a change in the thermal-electron 
distribution is not sufficient to explain these observations: they require 
that the direction of the line-of-sight component of the magnetic field in 
the region of Faraday rotation has changed between these two epochs. 
Since a change in the line-of-sight magnetic field is required, this makes
it natural to consider scenarios involving changes in a helical jet
magnetic field structure. 

Finally, we note that another possible way to explain time-variable
Faraday rotation is if compact jet components ``illuminate'' different
regions in a static, inhomogeneous foreground Faraday screen as they move
behind it, as was suggested by Zavala \& Taylor (2001) for the core RM 
variations in 3C279. However, this is likewise an unlikely explanation for the 
RM changes observed for B1803+784, because it is difficult to imagine how 
this effect could give rise to changes in the RM pattern {\em across} the 
jet. In addition, the apparent speeds of VLBI components near the region 
where we have detected the transverse RM gradients in B1803+784 are low: 
an essentially quasi-stationary component has been observed roughly 1.5~mas 
from the core for nearly three decades, and no features detected closer than
a few mas from the core show clear motions 
(Kellermann et al. 2004; Gabuzda \& Chernetskii 2003; 
http://www.physics.purdue.edu/MOJAVE/sepavstime/B1803 +784\_sepvstime.gif). 
This likewise appears to make it unlikely that the scenario envisaged for 
3C279 is operating in this region of the B1803+784 jet, since none of 
the observed compact VLBI features display appreciable motions.

\subsection{Propagation of the RM gradient reversal}

It is interesting to consider the possible motion or propagation of the 
transverse RM gradient pattern. To compare the distances from the core 
where the transverse RM gradient is most clearly visible, we must 
construct RM maps for the different epochs using the same restoring beam.
We have used the beam in Fig.~\ref{fig:apr1997jet} for this purpose,
and the RM maps for August 2002 and August 2003 constructed using this
beam are shown in Fig.~\ref{fig:same_beam}. We first note that 
the transverse gradients we initially identified based
on the images with their intrinsic beams are still visible in each of
these matched-resolution maps. This is particularly important for the
August 2002 RM image, which was made from data obtained at 15--43~GHz,
rather than 5--15~GHz, and so had an intrinsically appreciably higher
resolution. 

As in Fig.~\ref{fig:apr1997jet}, the transverse RM gradient is visible
essentially all along the VLBI structure, out to nearly 3~mas from the core. 
The distances from the core where the transverse RM gradient is most 
clearly visible in the remaining two matched resolution images are roughly 
$1.5-2.0$~mas in April 1997 and $2.0-2.5$~mas in August 2003; in the June 2000
RM map of Zavala \& Taylor (2003), the RM gradient is most prominent
$3-4$~mas from the core.    
(Note that we do not have access to the actual June 2000 RM map of Zavala \& 
Taylor (2003) in electronic form, and so our ability to analyze this image
and compare it with the other images considered is quite limited.)

The April 1997, June 2000 and 
August 2003 RM maps all have similar intrinsic resolutions, and so the 
positions where the transverse RM gradients are most clearly visible can be 
meaningfully compared.  However, we see no clear evidence for systematic 
movement of
the observed RM patterns either toward or away from the VLBI core over 
the time covered by these epochs: the distances from the core are 
$1.5-2.0$~mas in April 1997, $3-4$~mas in June 2000, and $2.0-2.5$~mas 
in August 2003. This is not entirely surprising, since the 
separation between our epochs may not be well matched to the time
scales on which the RM patterns evolve and/or propagate. For example, if the
RM pattern moves with a proper motion of several tenths
of a mas per year (typical of superluminal motions in AGN with redshifts
similar to B1803+784), it could move and evolve substantially over the 
three-year intervals from April 1997 to June 2000 and from June 2000 to 
August 2003, making it difficult to track the pattern. At the same time, 
the general
region of the transverse RM gradient is also near the location of
a long-lived quasi-stationary feature in the jet of B1803+784, approximately
1.5~mas from the VLBI core (e.g. Kellermann et al. 2004; Gabuzda \&
Chernetskii 2003). Depending on the origin of this quasi-stationary feature,
it may be that RM patterns in this region will also display very low
apparent speeds along the jet.  More frequent 
multi-frequency monitoring of this AGN with polarization VLBI could provide  
additional information about the evolution of the RM 
structure on parsec scales, which could potentially provide firmer
constraints on possible models for the origin of the observed RM gradient
and its evolution.

\section{Conclusion}
We have detected transverse RM gradients across the VLBI jet of B1803+784 at 
three different epochs; a transverse RM gradient is also visible in the 
RM distribution for this AGN published by Zavala \& Taylor (2003). The
presence of a transverse RM gradient across the jet at four different
epochs spanning about 7 years  provides firm evidence for a helical 
{\bf B} field associated with this jet.

Unexpectedly, a comparison of the RM gradients for these 4 epochs shows 
a clear {\em reversal} of the direction of the gradient between June 2000 
and August 2002. This is not accompanied by any obvious change in the
jet intensity or polarisation structure. 

The origin for this observed ``flip'' in the direction of 
the transverse RM gradient in the jet of B1803+784 is not clear. We have
suggested several scenarios that could give potentially rise to this phenomenon:
(i) a reversal of the direction of the poloidal component of the intrinsic 
magnetic field of the central
black hole; (ii) a reversal of the direction of the azimuthal field component
associated with torsional oscillations of the jet; and (iii) a change in
whether the ``inner'' or ``outer'' region of helical magnetic field dominates 
the total observed RM in a magnetic-tower picture. Although it is not
possible at this stage to conclusively identify whether any of these 
scenarios is the origin of the observed RM-gradient reversal, we favour 
the last hypothesis, because it provides a relatively simple explanation 
for this seemingly strange event,
since a ``nested helix'' structure for the jet magnetic field is a natural 
outcome of magnetic-tower models.

Further multi-frequency polarization studies of this source are clearly 
crucial for our understanding of how the transverse Faraday Rotation 
gradients evolve over time and with distance from the core. Higher resolution 
multi-frequency VLBI polarization studies can be used to study transverse 
RM gradients within the observed VLBI core in more detail, as is demonstrated
by our 15--43~GHz RM image. We have recently obtained further VLBA 
polarization observations of B1803+784 at 5--43~GHz, which we hope will 
enable us to trace the behaviour of the RM distribution over a wider range 
of distances along the jet, all at a single epoch. Numerical studies to 
determine the feasibility of various 
scenarios for the reversal of the transverse RM gradient would
also be of considerable interest, and are planned as part of our continuing
work in this area.

\section{Acknowledgements}
The research for this publication was financially supported by a Basic 
Research Grant from Science Foundation Ireland. The National Radio Astronomy 
Observatory is operated by Associated Universities Inc. We thank P. Cronin 
for his work on the RM map for April 1997, T. V. Cawthorne for useful
discussions of these results, and the anonymous referee for useful comments
and suggestions that have improved this paper.

\section{Appendix}

To illustrate why the presence of a RM gradient will not be affected
by incorrect EVPA calibration, we consider first the definition of the
RM, measured by comparing the polarisation angles $\chi_1$ and
$\chi_2$ measured at two wavelengths $\lambda_1$ and $\lambda_2$: 

\begin{equation}
RM = \frac{d\chi}{d\lambda^2} = \frac{\chi_2 - \chi_1}{\lambda_2^2 - \lambda_1^2}
\end{equation}

\noindent
The RM gradient between two points $a$ and $b$ is essentially the 
difference in the RM values measured at these two points:

\begin{eqnarray}
RM(a) - RM(b) & = & \frac{\chi_2(a) - \chi_1(a)}{\lambda_2^2 - \lambda_1^2} - \frac{\chi_2(b) - \chi_1(b)}{\lambda_2^2 - \lambda_1^2}\\
 & = & \frac{\chi_2(a) - \chi_2(b)}{\lambda_2^2 - \lambda_1^2} - \frac{\chi_1(a) - \chi_1(b)}{\lambda_2^2 - \lambda_1^2}
\end{eqnarray}

\noindent
We can see through the simple rearrangement in the second equation above that,
although uncertainty in the absolute EVPA calibration will contribute to the
absolute uncertainties of the RM values, the effects of incorrect EVPA 
calibration essentially cancel out when calculating the RM {\em gradient}, since
they will appear in the same way in the polarisation angles measured at
each point in the image at the wavelength in question, e.g., in $\chi_2(b)$ 
and $\chi_2(a)$. This fact must be taken into account when comparing RM
values at different points in an image and their uncertainties in order to
estimate the significance of observed RM gradients. For this reason, we have
estimated the uncertainties in the RM values at a given location in the map 
in Fig.~\ref{fig:trans_dist} purely from the rms deviation of the RM values in 
a $3\times 3$-pixel area 
surrounding this location, rather than using the formal uncertainties from the 
fits yielding the RM values, since the latter will be affected by possible
inaccuracies in the EVPA calibrations. 

\end{document}